\documentclass[journal, a4paper, comsoc]{IEEEtran}

\usepackage[utf8]{luainputenc}
\usepackage[noadjust]{cite}
 \usepackage{mathptmx}

\ifCLASSINFOpdf
\usepackage{graphicx}
\else
\fi

\usepackage[cmex10]{amsmath}
\usepackage{booktabs}
\usepackage{array}

\ifCLASSOPTIONcompsoc
  \usepackage[caption=false,font=normalsize,labelfont=sf,textfont=sf]{subfig}
\else
  \usepackage[caption=false,font=footnotesize]{subfig}
\fi

\usepackage{url}
\usepackage{amssymb, mathtools}
\usepackage{minamakron}
\graphicspath{{./illustrationer/}}
\let\oldmathbf\mathbf

\renewcommand{\mathbf}[1]{{\boldsymbol{\oldmathbf{#1}}}}

\usepackage{siunitx}
\usepackage{units}
\usepackage{microtype}

\usepackage{pgfplots}
\newlength\figureheight
\newlength\figurewidth 
\pgfplotsset{compat=newest}
\pgfplotsset{plot coordinates/math parser=false} 
\pgfplotsset{every x tick label/.append style={font=\small, yshift=0ex}}
\pgfplotsset{every y tick label/.append style={font=\small, xshift=0ex}}
\pgfplotsset{every axis legend/.append style={font=\small}}
\usetikzlibrary{calc}

\newtheorem{theorem}{Theorem}

\newtheorem{corollary}{Corollary}
\newtheorem{lemma}{Lemma}
\newtheorem{remark}{Remark}

\makeatletter
\tikzset{
	dot diameter/.store in=\dot@diameter,
	dot diameter=3pt,
	dot spacing/.store in=\dot@spacing,
	dot spacing=10pt,
	dots/.style={
		line width=\dot@diameter,
		line cap=round,
		dash pattern=on 0pt off \dot@spacing
	}
}
\makeatother

\usepackage{pdftexcmds}
\DeclareMathAlphabet{\mathsfbi}{OT1}{\sfdefault}{m}{sl}
\DeclareMathVersion{sfletters}
\SetSymbolFont{letters}{sfletters}{OML}{ntxsfmi}{m}{it}

\makeatletter
\newcommand{\mathbfsbilow}[1]{%
	\text{\mathversion{sfletters}$\m@th#1$}%
}

\DeclareRobustCommand{\mathsfit}[1]{%
	\begingroup
	\ifcat\noexpand #1\relax
	\edef\greek@test{\detokenize{#1}}%
	\edef\greek@test{\expandafter\@cdr\greek@test\@nil}%
	\edef\greek@test{\expandafter\@car\greek@test\@nil}%
	\edef\x{\the\lccode\expandafter`\greek@test}%
	\edef\y{\number\expandafter`\greek@test}%
	\ifnum\x=\y\relax
	\mathbfsbilow{#1}%
	\else
	\mathsfbi{#1}%
	\fi
	\else
	\mathsfbi{#1}%
	\fi
	\endgroup
}
\makeatother

\newcommand{\freq}[1]{\mathsfit{#1}}
\newcommand{\freqvec}[1]{\mathsf{#1}}

\renewcommand{\HTildeHatHermitian}{\mathbf{\freqvec{\hat{H}}}\llap{\phantom{$\mathbf{\freqvec{H}}$}}^\conjtr}
\renewcommand{\HTildeHat}{\mathbf{\freqvec{\hat{H}}}}
\renewcommand{\Exp}{\ensuremath{\mathbb{E}}}

\IEEEoverridecommandlockouts
\begin{document}
	\title{Uplink Performance of Wideband Massive~MIMO with One-Bit ADCs}
	
	\author{Christopher~Mollén, Junil~Choi, Erik~G.~Larsson, and~Robert~W.~Heath,~Jr.
		\thanks{C. Mollén and E. Larsson are with the Department of Electrical Engineering, Linköpings universitet, 581 83 Linköping, Sweden, e-mail: christopher.mollen@liu.se, erik.g.larsson@liu.se.}
		\thanks{J. Choi is with the Department of Electrical Engineering, Pohang University of Science and Technology (POSTECH), Pohang, Gyeongbuk, Korea 37673, e-mail: junil@postech.ac.kr.}%
		\thanks{R. Heath is with the Wireless Networking and Communications Group, The University of Texas at Austin, Austin, TX 78712, USA, e-mail: rheath@utexas.edu.}%
		\thanks{The research leading to these results has received funding from the European Union Seventh Framework Programme under grant agreement number \textsc{ict}-619086 (\textsc{mammoet}) and the Swedish Research Council (Vetenskapsrådet).}
		\thanks{This material is based upon work supported in part by the National Science Foundation under Grant No.\ \textsc{nsf-ccf-1527079}.}
		}
		
	\IEEEpubid{Accepted to IEEE Transactions on Wireless Communications.}
	
	\maketitle
	
	\begin{abstract}
		Analog-to-digital converters (\ADCs) stand for a significant part of the total power consumption in a massive \MIMO base station.  One-bit \ADCs are one way to reduce power consumption.  This paper presents an analysis of the spectral efficiency of single-carrier and \OFDM transmission in massive \MIMO systems that use one-bit \ADCs.  A closed-form achievable rate, i.e., a lower bound on capacity, is derived for a wideband system with a large number of channel taps that employs low-complexity linear channel estimation and symbol detection.  Quantization results in two types of error in the symbol detection.  The circularly symmetric error becomes Gaussian in massive \MIMO and vanishes as the number of antennas grows.  The amplitude distortion, which severely degrades the performance of \OFDM, is caused by variations between symbol durations in received interference energy.  As the number of channel taps grows, the amplitude distortion vanishes and \OFDM has the same performance as single-carrier transmission.  A main conclusion of this paper is that wideband massive \MIMO systems work well with one-bit \ADCs.
	\end{abstract}
	
	\begin{IEEEkeywords}
		channel estimation, equalization, \OFDM, one-bit \textsc{adc}s, wideband massive \MIMO.
	\end{IEEEkeywords}

	\section{Introduction}
	\IEEEPARstart{O}{ne-bit} Analog-to-Digital Converters (\ADCs) are the least power consuming devices to convert analog signals into digital \cite{walden1999analog}.  The use of one-bit \ADCs also simplifies the analog front end, e.g., automatic gain control (\AGC) becomes trivial because it only considers the sign of the input signal.  Such radically coarse quantization has been suggested for use in massive Multiple-Input Multiple-Output (\MIMO) base stations, where the large number of radio chains makes high resolution \ADCs a major power consumer.  Recent studies have shown that the performance loss due to the coarse quantization of one-bit \ADCs can be overcome with a large number of receive antennas \cite{bjornson2015massive, choi2016nearTCOM, jacobsson2016massive, Mo_Jianhua_TSP15}.
	
	Several recent papers have proposed specific symbol detection algorithms for massive \MIMO with low-resolution \ADCs.  For example, a near-maximum-likelihood detector for one-bit quantized signals was proposed in \cite{choi2016nearTCOM}, while \cite{jacobsson2016massive} and \cite{risi2014massive} studied the use of linear detection.  In \cite{liang2015mixed}, it was proposed to use a mix of one-bit and high resolution \ADCs, which was shown to give a performance similar to an unquantized system.  The proposed algorithms, however, focused only on frequency-flat channels. 
	
	Maximum-likelihood channel estimation for frequency-flat \MIMO channels with one-bit \ADCs was studied in \cite{ivrlac2007mimo}.  It was found that the quality of the estimates depends on the set of orthogonal pilot sequences used.  This is contrary to unquantized systems, where any set of orthogonal pilot sequences gives the same result.  Closed-loop channel estimation with 	dithering and “bursty” pilot sequences was proposed for the single-user frequency-selective channel in \cite{dabeer2010channel}.  It is not apparent that bursty pilot sequences are optimal and no closed-form expression for their performance was derived.  In \cite{mo2014channel, rusu2015adaptive, rusu2015low}, message passing algorithms were proposed that improve the estimation of sparse channels.  Our paper, in contrast, studies general non-sparse channels.  \IEEEpubidadjcol 
	
	Most previous work on massive \MIMO with one-bit \ADCs has focused on narrowband systems with frequency-flat channels, e.g.,\ \cite{choi2016nearTCOM, jacobsson2016massive, Mo_Jianhua_TSP15, risi2014massive, liang2015mixed, ivrlac2007mimo}.  Since quantization is a nonlinear operation on the time-domain signal, there is no straightforward way to extend these results to wideband systems, in which the channel is frequency selective.  Some recent work has proposed equalization and channel estimation algorithms for wideband systems \cite{studer2016quantized, wang2015multiuser, liang2016mixed}.  In \cite{studer2016quantized}, an iterative Orthogonal-Frequency-Division-Multiplexing (\OFDM) based equalization and channel estimation method was proposed.  However, the method is only shown to work for long pilot sequences of length $N_\text{d}K$ ($N_\text{d}$ the number of subcarriers, $K$ the number of users).  In contrast, our method only requires pilots of length $\mu KL$ ($L \ll N_\text{d}$ is the number of channel taps), where $\mu=1$ yields an acceptable performance in many cases.  Our method thus allows for a more efficient use of the coherence interval for actual data transmission.  In \cite{wang2015multiuser}, a message passing algorithm for equalization of single-carrier signals and a linear least-squares method for channel estimation were proposed.  The detection algorithm proposed in our paper is linear in the quantized signals and the channel estimation method is based on linear minimum-mean-square-error (\LMMSE) estimation, which has the advantage of performing relatively well independently of the noise variance.  The use of a mix of low and high-resolution \ADCs was also studied for frequency-selective single-user channels with perfect channel state information in \cite{liang2016mixed}.  However, the mixed \ADC architecture increases hardware complexity, in that an \ADC switch is required.  Furthermore, it is not clear that the computational complexity of the designs in \cite{studer2016quantized, wang2015multiuser, liang2016mixed} is low enough for real-time symbol detection, especially in wideband systems where the sampling rate is high.  
	
	In this paper, we study a massive \MIMO system with one-bit \ADCs and propose to apply to the quantized signals low-complexity linear combiners for symbol detection and \LMMSE channel estimation.  These are the same techniques that previously have been suggested for unquantized massive multiuser \MIMO \cite{marzetta2010noncooperative} and that have proven possible to implement in real time \cite{vieira2014flexible}.  Linear receivers for signals quantized by one-bit \ADCs have not been studied for frequency-selective channels before.
	
	Without any simplifying assumptions on the quantization distortion, we derive an achievable rate for single-carrier and \OFDM transmission in the proposed system, where the channel is estimated from pilot data and the symbols are detected with linear combiners.  A frequency-selective channel, in which the taps are Rayleigh fading and follow a general power delay profile, is assumed.  When the number of channel taps grows large, the achievable rate is derived in closed-form for the maximum-ratio and zero-forcing combiners (\MRC, \ZFC).  The rate analysis shows that simple linear receivers become feasible in wideband massive \MIMO systems, where the performance loss compared to an unquantized system is approximately \unit[4]{dB}.  In many system setups, the performance of the quantized system is approximately \unit[60--70]{\%} of the performance of the unquantized system at data rates around \unit[2]{bpcu}.  The loss can be made smaller, if longer pilot sequences can be afforded.  
	
	We also show that the performance of \OFDM, without any additional signal processing, is the same as the performance of single-carrier transmission in wideband systems with a large number of channel taps.  As was also noted in \cite{mollen2015onebit}, the quantization error of the symbol estimates consists of two parts: one amplitude distortion and one circularly symmetric distortion, whose distribution is close to Gaussian.  While the amplitude distortion causes significant intersymbol interference in \OFDM, it can easily be avoided in single-carrier transmission.  We show that the amplitude distortion vanishes when the number of taps grows.  Therefore only the circularly symmetric noise, which affects single-carrier and \OFDM transmission in the same way, is present in wideband systems with many taps.  Hence, frequency selectivity is beneficial for linear receivers in massive \MIMO because it reduces the quantization error and makes it circularly symmetric and additive.  Results in \cite{Mo_Jianhua_TSP15} indicate that the capacity of quantized \MIMO channels grows faster with the number of receive antennas at high signal-to-noise ratio (\SNR) than the rate of the linear combiners.  In a frequency-flat channel, where near-optimal detection becomes computationally feasible, a nonlinear symbol detection algorithm, e.g., \cite{choi2016nearTCOM, wang2015multiuser}, would therefore be better than linear detectors, especially at a high \SNR, where the linear \ZFC fails to suppress all interference in the quantized system, even with perfect channel state information.
		
	The most related work appeared in \cite{fan2015uplink, zhang2016spectral}, where achievable rates for massive \MIMO with one-bit \ADCs and low-resolution \ADCs for frequency-flat channels were investigated.  In \cite{fan2015uplink} an approximation was given for the achievable rate of a \MRC system with low-resolution \ADCs (one-bit \ADCs being a special case) for a Rayleigh fading channel.  The study showed a discrepancy between the approximation and the numerically obtained rate of one-bit \ADCs \cite[Figure~2]{fan2015uplink}, which was left unexplained.  In \cite{zhang2016spectral}, an approximation of an achievable rate for Ricean fading frequency-flat channels (of which Rayleigh fading is a special case) was derived.  Neither \cite{fan2015uplink} nor \cite{zhang2016spectral} mentioned that quantization distortion might combine coherently and result in amplitude distortion and neither mentioned the dependence of the rate on the number of channel taps.
	
	Parts of this work has been presented at \cite{mollen2016linear_WSA}, where the derivation of the achievable rate for \MRC, a special case, was outlined.  Channel estimation with fixed pilot lengths (equal to $KL$, i.e., with pilot excess factor 1) was also studied.  This paper is more general and complete: generic linear combiners are studied, the detailed derivation of the achievable rate is shown and the effects of different pilot lengths are analyzed.
	
	Paper disposition: The massive \MIMO uplink with one-bit \ADCs is presented in Section~\ref{sec:system_model}.  The quantization of one-bit \ADCs is studied in Section~\ref{sec:linear_model}.  Then the channel estimation is explained in Section~\ref{sec:channel_estimation}.  The uplink transmission is explained and analyzed in Section~\ref{sec:uplink_transmission}.  Finally, we present numerical results in Section~\ref{sec:numerical_examples} and draw our conclusion in Section~\ref{sec:conclusion}.  The program code used in the numerical part can be found at \cite{mollen2016onebitcode}.  
	
	\section{System Model}\label{sec:system_model}
\begin{figure*}
		\centering
		\scalebox{1}{\includegraphics[width=0.85\linewidth]{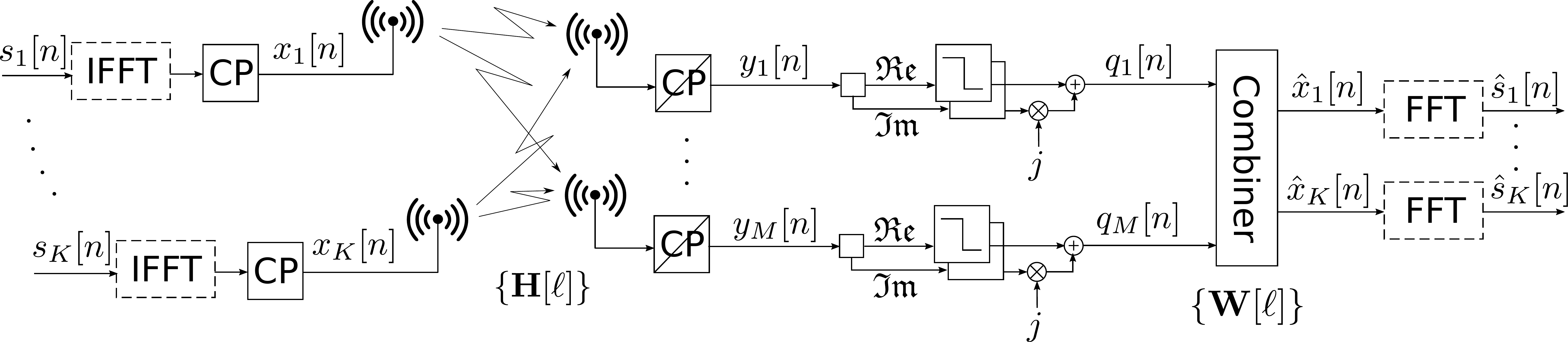}}
		\vspace{-0ex}
		\caption{The system model for the massive \MIMO uplink, both for single-carrier (without \textsc{ifft} and \textsc{fft}) and \OFDM transmission.}
		\label{fig:system_model}
	\end{figure*}

	We consider the massive \MIMO uplink in Figure~\ref{fig:system_model}, where the base station is equipped with $M$ antennas and there are $K$ single-antenna users.  All signals are modeled in complex baseband and are uniformly sampled at the Nyquist rate with perfect synchronization.  Because of these assumptions, the front-end depicted in Figure~\ref{fig:system_model} has been simplified accordingly.  Since the received signal is sampled at the Nyquist rate, there is no intermediate oversampled step before the matched receive filter (not in Figure~\ref{fig:system_model}), which thus has to be an analog filter.  Note that the one-bit \ADC itself does not require any \AGC---a dynamic control loop with variable attenuators and amplifiers that precisely adjusts the input voltage to conventional \ADCs to avoid clipping and to efficiently make use of the whole dynamic range that the \ADC has to offer.  Whereas one-bit \ADCs has no need for this kind of fine tuning of their input voltage, the analog receive filter, which probably would be an active filter, might require some kind of mechanism to regulate its input voltage to avoid being overdriven.  Such a mechanism would be simple to implement in comparison to the \AGC of a conventional \ADC and could possibly be combined with the low-noise amplifier \cite{kamgar2001lna}.
	
	At symbol duration $n$, base station antenna $m$ receives:
	\begin{align}\label{eq:rx_signal}
		y_m[n] \triangleq \! \sum_{k = 1}^{K} \sum_{\ell=0}^{L-1} \sqrt{P_k} g_{mk}[\ell] x_k[n - \ell] + z_m[n],
	\end{align}
	where $x_k[n]$ is the transmit signal from user $k$, whose power $\Exp\bigl[|x_k[n]|^2\bigr] = 1$, $P_k$ is the transmit power of user $k$ and $z_m[n] \sim \mathcal{CN}(0,N_0)$ is a random variable that models the thermal noise of the base station hardware.  It is assumed that $z_m[n]$ is identically and independently distributed (\IID) over $n$ and $m$ and independent of all other variables.  We assume that the $L$\mbox{-}tap impulse response $\{g_{mk}[\ell]\}$ of the channel between user $k$ and antenna $m$ can be written as the product of the small-scale fading $h_{mk}[\ell]$ and the large-scale fading $\sqrt{\beta_k}$:
	\begin{align}
		g_{mk}[\ell] = \sqrt{\beta_k} h_{mk}[\ell].
	\end{align}
	The small-scale fading has to be estimated by the base station.  Its mean $\Exp[h_{mk}[\ell]] = 0$ and variance are \textit{a priori} known:
	\begin{align}
		\Exp\bigl[ |h_{mk}[\ell]|^2 \bigr] = p[\ell], \quad \forall \ell,
	\end{align}
	where $p[\ell]$ is the power delay profile of the channel, for which
	\begin{align}
		\sum_{\ell=0}^{L-1} p[\ell] = 1.
	\end{align}
	In practice, the power delay profile depends on the propagation environment and has to be estimated, e.g., like in \cite{cui2006power}, where the power delay profile is estimated without additional pilots.  The base station is also assumed to know the large-scale fading $\beta_k$, which generally changes so slowly over time that an accurate estimate is easy to obtain in most cases.  The signal-to-noise ratio (\SNR) is defined as
	\begin{align}
		\SNR_k \triangleq \frac{P_k}{N_0} \sum_{\ell = 1}^{L-1} \Exp[|g_{mk}[\ell]|^2] = \frac{\beta_k P_k}{N_0}.
	\end{align}
	
	In a wideband system, the number of channel taps $L$ can be large---on the order of tens.  For example, a system that uses \unit[15]{MHz} of bandwidth over a channel with \unit[1]{\si{\micro\second}} of maximum excess delay, which corresponds to a moderately frequency-selective channel, has $L = 15$ channel taps.  The “Extended Typical Urban Model” \cite{3GPP_TS36.141_LTE_BS_testing} has a maximum excess delay of \unit[5]{\si{\micro\second}}, leading to $L = 75$~taps.
		
	Upon reception, the in-phase and quadrature signals are separately sampled, each by identical one-bit \ADCs, to produce:
	\begin{align}\label{eq:one_bit_quantization}
		q_m[n] \triangleq \frac{1}{\sqrt{2}} \sign\bigl(\Re (y_m[n]) \bigl) + j\frac{1}{\sqrt{2}} \sign\bigl(\Im (y_m[n]) \bigr).
	\end{align}
	We assume that the threshold of the quantization is zero.  Other thresholds can allow for better amplitude recovery when the interference and noise variance is small compared to the power of the desired signal \cite{narasimha2012ber, liang2015mixed}.  A small improvement in data rate can be obtained at low \SNR when a non-zero threshold is paired with the optimal symbol constellation, see \cite{koch2013low}, where the \SISO channel is studied.  Since we study a multiuser system, where the interuser interference is high, we do not expect any significant performance improvement from a non-zero threshold.  The scaling of the quantized signal is arbitrary but chosen such that $|q_m[n]| = 1$.  

	Two transmission modes are studied: single-carrier with frequency-domain equalization and \OFDM transmission.  We observe the transmission for a block of $N$ symbols.  At symbol duration $n$, user $k$ transmits
	\begin{align}
		x_k[n] = \begin{cases}
			s_k[n], &\text{if single-carrier}\\
			\! \frac{1}{\sqrt{N}} \! \sum_{\nu = 0}^{N-1} s_k[\nu] e^{j2\pi n\nu/N}, &\text{if \OFDM}
		\end{cases},
	\end{align}
	where $s_k[n]$ is the $n$\mbox{-}th data symbol.  We assume that the symbols have zero-mean and unit-power, i.e., $\Exp \bigl[s_k[n] \bigr] = 0$ and $\Exp \bigl[|s_k[n]|^2\bigr] = 1$ for all $k, n$.  The users also transmit a cyclic prefix that is $L-1$ symbols long:
	\begin{align}
		x_k[n] = x_k[N + n], \quad -L < n < 0,
	\end{align}
	so that the input-output relation in \eqref{eq:rx_signal} can be written as a multiplication in the frequency-domain, after the cyclic prefix has been discarded:
	\begin{align}\label{eq:freq_IO}
		\freq{y}_m[\nu] &= \sum_{k = 1}^{K} \sqrt{\beta_k P_k} \freq{h}_{mk}[\nu] \freq{x}_k[\nu] + \freq{z}_m[\nu],\\
	\intertext{where}
		\freq{x}_k[\nu] &\triangleq \frac{1}{\sqrt{N}} \sum_{n=0}^{N-1} x_k[n] e^{-j2 \pi n\nu/N}\label{eq:90123780918}\\
		\freq{y}_m[\nu] &\triangleq \frac{1}{\sqrt{N}} \sum_{n=0}^{N-1} y_m[n] e^{-j2 \pi n\nu / N}\label{eq:92038019}\\
		\freq{h}_{mk}[\nu] &\triangleq \sum_{\ell = 0}^{L-1} h_{mk}[\ell] e^{-j2 \pi \ell \nu / N}\label{eq:923047990}
	\end{align}
	and $\freq{z}_m[\nu] \sim \mathcal{CN}(0, N_0)$ \IID is the unitary Fourier transform of the thermal noise.  

\section{Quantization}\label{sec:linear_model}
In this section, some properties of the quantization of one-bit \ADCs are derived.  These results are used later in the channel estimation and the system analysis.  

We define the quantization distortion as
\begin{align}\label{eq:quantization_noise}
	e_m[n] \triangleq q_m[n] - \rho y_m[n],
\end{align}
where the scaling factor $\rho$ is chosen to minimize the error variance $E \triangleq \Exp\bigl[|e_m[n]|^2\bigr]$, which is minimized by the Wiener solution:
\begin{align}
	\rho = \frac{\Exp \bigl[y^*_m[n] q_m[n] \bigr]}{\Exp\bigl[ |y_m[n]|^2 \bigr]}.\label{eq:expression_scaling_parameter}
\end{align}

Note that the distribution of $e_m[n]$ depends on the distribution of the received signal $y_m[n]$ in a nonlinear way due to \eqref{eq:one_bit_quantization} and that $e_m[n]$ is uncorrelated to $y_m[n]$ due to the choice of $\rho$ because of the orthogonality principle.  
	
	We define the expected received power given all transmit signals:
	\begin{align}
		P_\text{rx}[n] &\triangleq \Exp\big[ |y_m[n]|^2 \bigm| \{x_k[n]\} \big] \notag\\
		&= N_0 + \sum_{k = 1}^{K} \beta_k P_k \sum_{\ell=0}^{L-1} p[\ell] \big| x_k[n - \ell] \big|^2,\label{eq:912347423}\\
	\intertext{and the average received power:}
		\bar{P}_\text{rx} &\triangleq \Exp\big[ |y_m[n]|^2 \big] = N_0 + \sum_{k = 1}^{K} \beta_k P_k.\label{eq:816727511}
	\end{align}
	When the number of channel taps $L$ in \eqref{eq:912347423} is large, the two powers $P_\text{rx}[n]$ and $\bar{P}_\text{rx}$ are close to equal.  This is formalized in the following lemma.
	\begin{lemma}\label{lem:widebandMIMOapprox}
		Given a sequence of increasingly long power delay profiles $\{p_L[\ell]\}_{L=1}^\infty$ and a constant $\gamma \in \mathbb{R}$ such that $\max_\ell p_L[\ell] < \gamma / L$, for all lengths $L$, then
		\begin{align}\label{eq:817973391}
			P_\text{rx}[n] \xrightarrow{\text{a.s.}} \bar{P}_\text{rx}, \quad L\to\infty, \qquad \forall n.
		\end{align}
	\end{lemma}
	\begin{IEEEproof}
		According to the law of large numbers and the Kolmogorov criterion \cite[eq.~7.2]{feller1968introduction},
		\begin{align}
			\sum_{\ell=0}^{L-1} p_L[\ell] \bigl|x_k[n-\ell]\bigr|^2 \xrightarrow{\text{a.s.}} \Exp\Bigl[ \bigl|x_k[n-\ell]\bigr|^2 \Bigr] = 1, \quad L \to \infty.
		\end{align}
		Thus, the inner sum in \eqref{eq:912347423} tends to one as the number of channel taps grows. 
	\end{IEEEproof}
	
	Because of the cyclic prefix, the block length $N$ cannot be shorter than $L$.  Therefore, it was assumed that $N$ grew together with $L$ in the proof of Lemma~\ref{lem:widebandMIMOapprox}.  As we will see later, the convergence can be fast, so the left-hand side in \eqref{eq:817973391} is well approximated by its limit also for $L$ much shorter than usual block lengths $N$.  
	
	\begin{remark}\label{rem:big_K}
		Note that the sum over $k$ in \eqref{eq:912347423} also results in an averaging effect when the number of users $K$ is large.  The relative difference between the expected received power given all transmit signals and its mean \mbox{$|P_\text{rx}[n] - \bar{P}_\text{rx}| / P_\text{rx}[n]$} becomes small, not only with increasing $L$, but also with increasing $K$ if there is no dominating user, i.e., some user $k$ for which $\beta_k P_k \gg \beta_{k'} P_{k'}, \forall k' \neq k$.  In practice, power control is done and most $\beta_k P_k$ will be of similar magnitude.  The expected received power given all transmit signals is thus closely approximated by its average also in narrowband systems with a large number of users and no dominating user.
	\end{remark}
	
	The next lemma gives the scaling factor and the variance of the quantization distortion.
	\begin{lemma}\label{lem:rho_and_error_var}
		If the fading is \IID Rayleigh, i.e., $h_{mk}[\ell] \sim \mathcal{CN}(0, p[\ell])$ for all $m$, $k$ and $\ell$, then the scaling factor defined in \eqref{eq:expression_scaling_parameter} is given by
		\begin{align}
			\rho &= \sqrt{\frac{2}{\pi}} \frac{\Exp \Bigl[ \sqrt{P_\text{rx}[n]} \Bigr]}{\bar{P}_\text{rx}} \to \sqrt{\frac{2}{\pi \bar{P}_\text{rx}}}, \quad L\to\infty,\label{eq:rho}\\
		\intertext{and the quantization distortion has the variance}
			E &= 1 - \rho^2 \bar{P}_\text{rx} \to 1 - \frac{2}{\pi}, \quad L \to \infty.\label{eq:error_variance}
		\end{align}
	\end{lemma}
\begin{IEEEproof}
	See Appendix~\ref{app:rho_and_error_var}.
\end{IEEEproof}

	We see that the error variance in \eqref{eq:error_variance} would equal its limit if $P_\text{rx}[n] = \bar{P}_\text{rx}$ and $\rho^2 = \frac{2}{\pi \bar{P}_\text{rx}}$.  That is the reason the limit coincides with the mean-squared error of one-bit quantization in \cite{max1960quantizing} and what is called the distortion factor of one-bit \ADCs in \cite{bai2015energy}.

	The following corollary to Lemma~\ref{lem:rho_and_error_var} gives the limit of the relative quantization distortion variance, which is defined as
	\begin{align}
	Q \triangleq \frac{E}{|\rho|^2}.
	\end{align}
	\begin{corollary}\label{the:noise_power}
		The relative quantization distortion variance in a wideband system approaches
		\begin{align}\label{eq:one_bit_quantization_power}
			Q \to Q' \triangleq \bar{P}_\text{rx} \Big(\frac{\pi}{2} - 1 \Big), \quad L\to\infty.
		\end{align}
	\end{corollary}

Note that $Q \geq Q'$, because Jensen's inequality says that $\rho \leq \sqrt{\frac{2}{\pi \bar{P}_\text{rx}}}$ is smaller than its limit in \eqref{eq:rho} for all $L$, since the square root is concave.  This means that the variance of the quantization distortion is smaller in a wideband system, where the number of taps $L$ is large, than in a narrowband system.

If there is no quantization, the variance of the quantization error $E = 0$ and thus the relative quantization distortion $Q = Q' = 0$.  This allows us to use the expressions derived in the following sections to analyze the unquantized system as a special case.
	
\section{Channel Estimation}\label{sec:channel_estimation}
In this section, we will describe a low-complexity channel estimation method.  In doing so, we assume that the uplink transmission is divided into two blocks: one with length $N = N_\text{p}$ pilot symbols for channel estimation and one with length $N = N_\text{d}$ symbols for data transmission.  The two blocks are disjoint in time and studied separately.  It is assumed, however, that the channel is the same for both blocks, i.e., that the channel is block fading and that both blocks fit within the same coherence time.

We define $K$ orthogonal pilot sequences of length $N_\text{p}$ as:
\begin{align}
	\freq{\phi}_k[\nu] \triangleq \begin{cases}
	0, &(\nu \!\mod{K}) + 1 \neq k\\
	\sqrt{\frac{K}{N_\text{p}}}e^{j\theta_k[\nu]}, &(\nu \!\mod{K}) + 1 = k
	\end{cases},
\end{align}
where $\theta_k[\nu]$ is a phase that is known to the base station.  During the training period, user $k$ transmits the signal that, in the frequency domain, is given by
\begin{align}
	\freq{x}_k[\nu] = \sqrt{N_\text{p}} \freq{\phi}_k[\nu].
\end{align}
The received signal \eqref{eq:freq_IO} is then
\begin{align}
	\freq{y}_m[\nu] 
	&= \sum_{k = 1}^{K} \sqrt{\beta_{k} P_{k} N_\text{p}} \freq{h}_{mk}[\nu] \freq{\phi}_{k}[\nu] + \freq{z}_m[\nu]\\
	&= \sqrt{\beta_{k'} P_{k'} K} \freq{h}_{mk'}[\nu] e^{j\theta_{k'}[\nu]} + \freq{z}_m[\nu],\label{eq:98712304789}
\end{align}
where $k' \triangleq (\nu \mod{K}) + 1$, in the last step, is the index of the user whose pilot $\freq{\phi}_{k'}[\nu]$ is nonzero at tone $\nu$.  By rewriting the time-domain quantized signal using \eqref{eq:quantization_noise}, we compute the quantized received signal in the frequency domain as
\begin{align}
	\freq{q}_m[\nu] &\triangleq \frac{1}{\sqrt{N_\text{p}}} \sum_{n = 0}^{N_\text{p} - 1} q_m[n] e^{-j 2 \pi n \nu / N_\text{p}}\\
	&= \rho \freq{y}_m[\nu] + \underbrace{\frac{1}{\sqrt{N_\text{p}}} \sum_{n = 0}^{N_\text{p} - 1} e_m[n] e^{-j 2 \pi n \nu / N_\text{p}}}_{\triangleq \freq{e}_m[\nu]}\\
	&= \rho \sqrt{\beta_{k'} P_{k'} K} \freq{h}_{mk'}[\nu] e^{j\theta_{k'}[\nu]} + \rho \freq{z}_m[\nu] + \freq{e}_m[\nu].\label{eq:channel_est_freq_sample}
\end{align}
The sequence $\{\freq{q}_m[\nu K + k - 1], \, \nu = 0, \ldots, \frac{N_p}{K} - 1\}$ is thus a phase-rotated and noisy version of the frequency-domain channel of user $k$, sampled with period $F = K$.  

Because the time-domain channel $h_{mk}[\ell] = 0$ for all $\ell \notin [0,L-1]$, the sampling theorem says that, if the sampling period satisfies
\begin{align}\label{eq:sampling_theorem_req}
	F \leq \frac{N_\text{p}}{L},
\end{align}
it is enough to know the samples $\{\freq{h}_{mk}[\nu F + k - 1], \nu = 0, \ldots, \frac{N_\text{p}}{F} - 1\}$ of the channel spectrum to recover the time-domain channel:
\begin{align}\label{eq:channel_recovery}
	h_{mk}[\ell] = \frac{F}{N_\text{p}} \sum_{n = 0}^{\frac{N\text{p}}{F} - 1} \freq{h}_{mk}[nF + k -1] e^{j 2 \pi \ell (n F + k - 1) / N_\text{p}}.
\end{align}
Thus, if the number of pilot symbols satisfies $N_\text{p} \geq KL$, then \eqref{eq:sampling_theorem_req} is fulfilled and the following observation of the channel tap $h_{mk}[\ell]$ can be made through an inverse Fourier transform of the received samples that belong to user $k$:
\begin{align}
	&h'_{mk}[\ell] \notag\\
	&\triangleq \sqrt{\frac{K}{N_\text{p}}} \sum_{\nu = 0}^{\frac{N_\text{p}}{K} - 1} \freq{q}_m[\nu K + k - 1] e^{j 2 \pi \ell (\nu K + k - 1) / N_\text{p}} e^{-j\theta_k[\nu K + k - 1]}\label{eq:190872340}\\
	&= \rho \sqrt{\beta_k P_k K} \sqrt{\frac{K}{N_\text{p}}} \sum_{\nu = 0}^{\frac{N_\text{p}}{K} - 1} \freq{h}_{mk}[\nu K {+} k {-} 1] e^{j 2 \pi \ell (\nu K + k - 1) / N_\text{p}} \notag\\
	&\quad+ \rho \underbrace{\sqrt{\frac{K}{N_\text{p}}} \sum_{\nu = 0}^{\frac{N_\text{p}}{K} - 1} \freq{z}_{m}[\nu K {+} k {-} 1] e^{j 2 \pi \ell (\nu K + k - 1) / N_\text{p}} e^{-j\theta_k[\nu K + k - 1]}}_{\triangleq z'_{mk}[\ell]} \notag\\
	&\quad+ \underbrace{\sqrt{\frac{K}{N_\text{p}}} \sum_{\nu = 0}^{\frac{N_\text{p}}{K} - 1} \freq{e}_m[\nu K {+} k {-} 1] e^{j 2 \pi \ell (\nu K + k - 1) / N_\text{p}} e^{-j\theta_k[\nu K + k - 1]}}_{\triangleq e'_{mk}[\ell]}\label{eq:81828888}\\
	&= \rho \sqrt{\beta_k P_k N_\text{p}} h_{mk}[\ell] + \rho z'_{mk}[\ell] + e'_{mk}[\ell].\label{eq:929838383}
\end{align}
In the first step \eqref{eq:81828888}, $\freq{q}_m[\nu]$ is replaced by the expression in \eqref{eq:channel_est_freq_sample}.  Then, in \eqref{eq:929838383}, the relation in \eqref{eq:channel_recovery} is used to rewrite the first sum as the time-domain channel impulse response.  We note that the Fourier transform is unitary and therefore $z'_{mk}[\ell] \sim \mathcal{CN}(0, N_0)$ is independent across $m, k, \ell$ and $\Exp\bigl[ |e'_{mk}[\ell]|^2 \bigr] = \Exp \bigl[ |\freq{e}'_{mk}[\ell]|^2 \bigr] = E$.

We use the \LMMSE estimate of the channel, which is given by
\begin{align}
	\hat{h}_{mk}[\ell] &\triangleq h'_{mk}[\ell] \frac{\Exp \bigl[ h^*_{mk}[\ell] h'_{mk}[\ell]\big]^*}{\Exp\bigl[ |h'_{mk}[\ell]|^2 \bigr]}\\
	&= h'_{mk}[\ell] \frac{\rho p[\ell] \sqrt{\beta_k P_k N_\text{p}} }{\rho^2 p[\ell] \beta_k P_k N_\text{p} + \rho^2 N_0 + E}
\end{align}
and whose variance is $\Exp \bigl[ |\hat{h}_{mk}[\ell]|^2 \bigr] =  c_k[\ell] p[\ell]$, where
\begin{align}
	c_k[\ell] &\triangleq \frac{p[\ell] \beta_k P_k N_\text{p}}{p[\ell] \beta_k P_k N_\text{p} + N_0 + Q}.\label{eq:channel_powerloss}
\end{align}
The error $\epsilon_{mk}[\ell] \triangleq h_{mk}[\ell] - \hat{h}_{mk}[\ell]$ is uncorrelated to the channel estimate and has variance
\begin{align}\label{eq:838383838}
	\Exp\bigl[ | \epsilon_{mk}[\ell] |^2 \bigr] = \bigl( 1 - c_k[\ell] \bigr) p[\ell].
\end{align}

In the frequency domain, the channel estimate is given by
\begin{align}\label{eq:8189191}
	\freq{\hat{h}}_{mk}[\nu] \triangleq \sum_{\ell=0}^{L-1} \hat{h}_{mk}[\ell] e^{-j2\pi \nu \ell / N_\text{d}}, \quad \nu = 0, \ldots, N_\text{d} - 1,
\end{align}
the estimation error $\freq{\varepsilon}_{mk}[\nu] \triangleq \freq{h}_{mk}[\nu] - \freq{\hat{h}}_{mk}[\nu]$ and their variances:
\begin{align}
	\Exp \bigl[| \freq{\hat{h}}_{mk}[\nu] |^2\bigr] &= \sum_{\ell=0}^{L-1} c_k[\ell] p[\ell] \triangleq c_k\\
	\Exp \bigl[|\freq{\varepsilon}_{mk}[\nu]|^2\bigr] &= 1 - c_k.\label{eq:channel_error_var}
\end{align}
The variance of the channel estimate $c_k$ will be referred to as the channel estimation quality.  

We define the pilot excess factor as $\mu \triangleq \frac{N_\text{p}}{KL} \geq 1$.  Because $c_k \to 1$ as $\mu \to \infty$, the quality of the channel estimation in the quantized system can be made arbitrary good by increasing $\mu$.  Since the pilots have to fit within the finite coherence time of the channel however, the channel estimation quality will be limited in practice.  To get a feeling for how large practical pilot excess factors can be, we consider an outdoor channel with Doppler spread $\sigma_\nu = \unit[400]{Hz}$ and delay spread $\sigma_\tau = \unit[3]{\si{\micro\second}}$ and symbol duration $T$.  The coherence time of this channel is approximately $N_\text{c} \approx 1/(\sigma_\nu T)$ symbol durations and the number of taps $L \approx \sigma_\tau/T$.  The pilots sequence will thus fit if $N_\text{p} \leq N_\text{c}$, i.e., only pilot excess factors such that $\mu \leq 1/(K \sigma_\nu \sigma_\tau) \approx 830 / K$ are feasible in this channel.  Because of the finite coherence time of the practical channel, we will study the general case of finite $\mu$.

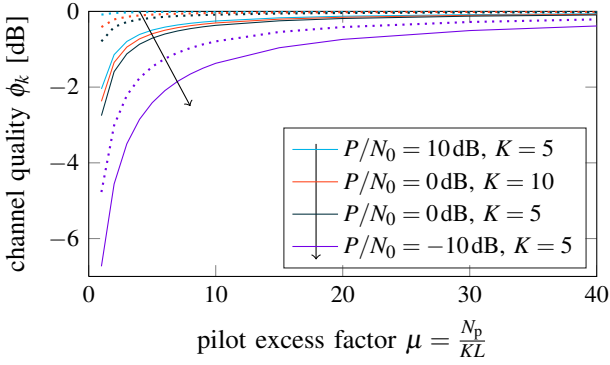
\begin{figure}
	\setlength{\figurewidth}{20em}
	\setlength{\figureheight}{10em}
	\centering
	\scalebox{1}{

\definecolor{mycolor1}{rgb}{0.00000,0.72266,0.90234}%
\definecolor{mycolor2}{rgb}{1.00000,0.39216,0.25882}%
\definecolor{mycolor3}{HTML}{8781d3}%
\definecolor{mycolor4}{HTML}{687F91}
\begin{tikzpicture}

\begin{axis}[%
width=0.951\figurewidth,
height=\figureheight,
at={(0\figurewidth,0\figureheight)},
scale only axis,
xmin=0,
xmax=40,
xlabel={pilot excess factor $\mu = \frac{N_\text{p}}{KL}$},
ymin=-7, 
ymax=0,
ylabel={channel quality $\phi_k$ [dB]},
axis background/.style={fill=white},
legend style={at={(.98,.03)}, anchor=south east,legend cell align=left,align=left,draw=white!15!black, fill = none}
]

\addplot [color=color1,solid]
table[row sep=crcr]{%
	1	-2.0472004879207\\
	2	-1.14312721884726\\
	3	-0.794480444730413\\
	4	-0.609066530621639\\
	5	-0.493895547577549\\
	6	-0.415382318900444\\
	7	-0.358419508991688\\
	8	-0.315201672588053\\
	9	-0.281287922179331\\
	10	-0.253964944317782\\
	15	-0.170949227152346\\
	20	-0.128838614929811\\
	30	-0.08631568320283\\
	40	-0.0648970587581807\\
};
\addlegendentry{$P/N_0 = \unit[10]{\text{dB}}$, $K = 5$};

\addplot [color=color1,dotted,line width = .8pt,forget plot]
table[row sep=crcr]{%
	1	-0.0860017176191763\\
	2	-0.0432137378264258\\
	3	-0.0288568823748821\\
	4	-0.0216606175650762\\
	5	-0.0173371280900053\\
	6	-0.0144524087418088\\
	7	-0.0123907211554842\\
	8	-0.010843812922199\\
	9	-0.00964028102616848\\
	10	-0.00867721531226935\\
	15	-0.00578673612468286\\
	20	-0.00434077479318578\\
	30	-0.00289433187589164\\
	40	-0.00217092972230206\\
};

\addplot [color=color2,solid]
table[row sep=crcr]{%
	1	-2.37512562188378\\
	2	-1.34794622761602\\
	3	-0.943401985835184\\
	4	-0.726060825472525\\
	5	-0.590236001629603\\
	6	-0.497267254780908\\
	7	-0.429621025665574\\
	8	-0.37818578441389\\
	9	-0.3377551481877\\
	10	-0.305137433644932\\
	15	-0.205788163097805\\
	20	-0.155248054003855\\
	30	-0.104112863059636\\
	40	-0.0783177001918832\\
};
\addlegendentry{$P/N_0 = \unit[0]{\text{dB}}$, $K = 10$};

\addplot [color=color2,dotted, line width = .8pt,forget plot]
table[row sep=crcr]{%
	1	-0.413926851582251\\
	2	-0.211892990699381\\
	3	-0.142404391146101\\
	4	-0.10723865391773\\
	5	-0.0860017176191763\\
	6	-0.0717858462712341\\
	7	-0.0616030870481854\\
	8	-0.0539503188670609\\
	9	-0.0479888288176871\\
	10	-0.0432137378264248\\
	15	-0.0288568823748821\\
	20	-0.0216606175650762\\
	30	-0.0144524087418088\\
	40	-0.010843812922199\\
};

\addplot [color=color3,solid]
table[row sep=crcr]{%
	1	-2.75301123077777\\
	2	-1.59109136825704\\
	3	-1.12264804146292\\
	4	-0.868006286878966\\
	5	-0.707731856717625\\
	6	-0.597498930263326\\
	7	-0.517012117233554\\
	8	-0.455652745100158\\
	9	-0.407322131607106\\
	10	-0.368266615698677\\
	15	-0.248947459992393\\
	20	-0.188035604943516\\
	30	-0.126257265054055\\
	40	-0.0950353904454516\\
};
\addlegendentry{$P/N_0 = \unit[0]{\text{dB}}$, $K = 5$};

\addplot [color=color3,dotted,line width = .8pt,forget plot]
table[row sep=crcr]{%
	1	-0.791812460476248\\
	2	-0.413926851582251\\
	3	-0.280287236002435\\
	4	-0.211892990699381\\
	5	-0.170333392987804\\
	6	-0.142404391146102\\
	7	-0.122344564170117\\
	8	-0.107238653917731\\
	9	-0.0954531790623039\\
	10	-0.0860017176191763\\
	15	-0.0575232888909132\\
	20	-0.0432137378264258\\
	30	-0.0288568823748821\\
	40	-0.0216606175650762\\
};

\addplot [color=color4,solid]
table[row sep=crcr]{%
	1	-6.73241131749815\\
	2	-4.5578777703321\\
	3	-3.4975586103003\\
	4	-2.85128934101882\\
	5	-2.41167252708238\\
	6	-2.09174817154349\\
	7	-1.84788294057613\\
	8	-1.65555500324611\\
	9	-1.49984663734989\\
	10	-1.37113124418575\\
	15	-0.960379776568521\\
	20	-0.739453149035155\\
	30	-0.506682745511272\\
	40	-0.385445509806447\\
};
\addlegendentry{$P/N_0 = \unit[-10]{\text{dB}}$, $K = 5$};

\addplot [color=color4,dotted,line width = .8pt,forget plot]
table[row sep=crcr]{%
	1	-4.77121254719663\\
	2	-3.01029995663981\\
	3	-2.21848749616356\\
	4	-1.76091259055681\\
	5	-1.46128035678238\\
	6	-1.249387366083\\
	7	-1.09144469425068\\
	8	-0.969100130080564\\
	9	-0.871501757189002\\
	10	-0.791812460476248\\
	15	-0.543576623225926\\
	20	-0.413926851582251\\
	30	-0.280287236002435\\
	40	-0.211892990699381\\
};

\draw[->] (axis cs:4,0) -- (axis cs: 8,-2.5); 
\draw[->] (axis cs:17.9,-3.5) -- (axis cs:17.9,-6.55); 

\end{axis}
\end{tikzpicture}%
}
	\vspace{-0ex}
	\caption{The channel estimation quality for the quantized system with one-bit \ADCs in solid lines and for the unquantized system in dotted lines for a uniform power delay profile $p[\ell] = 1 / L$.}
	\label{fig:channel_estimation_powerloss}
\end{figure}

To compare the channel estimation quality of a quantized wideband system to that of the corresponding unquantized system, we define 
\begin{align}
	c_k\bigr|_{Q=0, \mu = \mu_0} &\triangleq \sum_{\ell=0}^{L-1} \frac{p^2[\ell] \beta_k P_k \mu_0 K L}{p[\ell] \beta_k P_k \mu_0 K L + N_0}\\
	c_k\bigr|_{Q = \bar{P}_\text{rx}(\frac{\pi}{2} - 1), \mu = \mu_\text{q}} &\triangleq \sum_{\ell=0}^{L-1} \frac{p^2[\ell] \beta_k P_k \mu_\text{q} K L}{p[\ell] \beta_k P_k \mu_\text{q} K L + N_0 + \bar{P}_\text{rx}(\frac{\pi}{2} - 1)}\\
	\Delta(\mu_0, \mu_\text{q}) &\triangleq \frac{c_k \bigr|_{Q=0, \mu=\mu_0}}{c_k\bigr|_{Q = \bar{P}_\text{rx}(\frac{\pi}{2} - 1), \mu = \mu_\text{q}}},
\end{align}
where $\mu_0$ is the excess factor of the unquantized system, $\mu_\text{q}$ is that of the one-bit \ADC system and $\Delta(\mu_0, \mu_\text{q})$ the quality ratio.  If this ratio is one, the channel estimates of the two systems are equally good.  Under the assumption that $\beta_k P_k = P$, $\forall k$, and $p[\ell] = 1 / L$, $\forall \ell$, the difference in channel estimation quality always is less than \unit[2]{dB} when the excess factors $\mu_\text{q} = \mu_0 = \mu$ are the same:
\begin{align}\label{eq:9982341}
	\Delta(\mu, \mu) \leq \frac{\pi}{2} \approx \unit[2]{\text{dB}},
\end{align}
Further, it can be seen that $\Delta(1, 1) = \pi / 2$ and that $\Delta(\mu, \mu)$ is decreasing in $\mu$. This can be seen in Figure~\ref{fig:channel_estimation_powerloss}, where the channel estimation quality $c_k$ is plotted for different \SNR{}s $P / N_0$, where $\beta_k P_k = P$ for all $k$.  It can be seen that the power loss due to channel estimation is small in many system setups---in the order of \unit[$-$2]{dB}.

\begin{remark}
	To increase the length of the training period and to increase the transmit power of the pilot signal would give the same improvement in channel estimation quality in the unquantized system.  Because the orthogonality of the pilots is broken by the quantization, this is not true for the quantized system, which can be seen in \eqref{eq:channel_powerloss}, where $Q$ is a function of only the transmit power.  This is the reason the phases $\theta_k[n]$ are introduced: non-constant phases allow for pilot excess factors $\mu > 1$.  Note that, with constant phases (assume $\theta_k[n] = 0$ without loss of generality), the pilot signal transmitted during the training period \eqref{eq:98712304789} is sparse in the time domain:
	\begin{align}
		x_k[n] = 
		\begin{cases}
			\sqrt{\mu L}, &\text{if } n = \nu\mu L + k - 1, \quad \nu \in \mathbb{Z}\\
			0, &\text{otherwise}
		\end{cases},
	\end{align}
	i.e., it is zero in intervals of width $\mu L - 1$.  If $\mu > 1$ there are intervals, in which nothing is received.  The estimate is then based on few observations, each with relatively high \SNR.  By choosing the phases such that they are no longer constant, for example according to a uniform distribution $\theta_k[\nu] \sim \operatorname{unif}[0, 2 \pi)$, the pilot signal is no longer sparse in the time domain.  The estimate is then based on many observations, each with relatively low \SNR.  Increasing the number of low-\SNR observations is a better way to improve the quality of the channel estimate than increasing the \SNR of a few observations in a quantized system.  Furthermore, the limits in Lemma~\ref{lem:rho_and_error_var} are only valid if the received power $P_\text{rx}[n]$ becomes constant as $L \to \infty$, which is not the case when there are intervals, in which nothing is received.  
\end{remark}

\section{Uplink Data Transmission}\label{sec:uplink_transmission}
In this section, one block of $N = N_\text{d}$ symbols of the uplink data transmission is studied.  Practical linear symbol detection based on the estimated channel is presented and applied to the massive \MIMO system with one-bit \ADCs.  The distribution of the symbol estimation error due to quantization and how \OFDM is affected is also analyzed.  Finally, the performance is evaluated by deriving an achievable rate for the system.

\subsection{Receive Combining}
Upon reception, the base station combines the received signals using an \textsc{fir} filter with transfer function $\mathsfit{w}_{km}[\nu]$ and impulse response
\begin{align}\label{eq:combiner_impulse_response}
w_{km}[\ell] \triangleq \frac{1}{N_\text{d}} \sum_{\nu = 0}^{N_\text{d}-1} \mathsfit{w}_{km}[\nu] e^{j2\pi \nu\ell/N_\text{d}}, \quad \ell = 0, \ldots, N_\text{d} - 1
\end{align}
to obtain an estimate of time-domain transmit signals:
\begin{align}\label{eq:RxC}
	\hat{x}_k[n] \triangleq \sum_{m=1}^{M} \sum_{\ell=0}^{N_\text{d} - 1} w_{km}[\ell] q_m\bigl[ [n - \ell]_{N_\text{d}} \bigr],
\end{align}
where $[n]_{N_\text{d}} \triangleq n \mod{N_\text{d}}$, and, equivalently, of the frequency-domain transmit signals
\begin{align}\label{eq:RxC_freq}
	\freq{\hat{x}}_k[\nu] \triangleq \sum_{m=1}^{M} \freq{w}_{km}[\nu] \freq{q}_m[\nu],
\end{align}
where $\freq{q}_m[\nu]$ is the Fourier transform of the quantized signals.  The symbol estimate of user $k$ is then obtained as
\begin{align}\label{eq:symbol_estimate}
\hat{s}_k[n] = \begin{cases}
\hat{x}_k[n], &\text{ if single-carrier}\\
\freq{\hat{x}}_k[n], &\text{ if \OFDM}
\end{cases}.
\end{align}
The combiner weights are derived from the estimated channel matrix $\boldsymbol{\freqvec{\hat{H}}}[\nu]$, whose $(m,k)$\mbox{-}th element is $\freq{\hat{h}}_{mk}[\nu]$.  Three common combiners are the Maximum-Ratio, Zero-Forcing, and Regularized Zero-Forcing Combiners (\MRC, \ZFC, \RZFC):
\begin{align}\label{eq:combiners}
	\mathbf{\freqvec{W}}\llap{\phantom{W}}'[\nu] = \begin{cases}
		\HTildeHatHermitian[\nu], &\text{ if \textsc{mrc}}\\
		\bigl( \HTildeHatHermitian[\nu] \HTildeHat[\nu] \bigr)^{-1} \HTildeHatHermitian[\nu], 
		&\text{ if \textsc{zfc}}\\
		\bigl(\HTildeHatHermitian[\nu] \HTildeHat[\nu] + \lambda\mathbf{I}_K \bigr)^{-1} \HTildeHatHermitian[\nu], &\text{ if \textsc{rzfc}}
	\end{cases},
\end{align}
where $\lambda$ is a regularization factor.  The energy scaling of the combiner weights is arbitrary; for convenience, it is chosen as follows:
\begin{align}
	\freq{w}_{km}[\nu] = \frac{1}{\sqrt{\alpha_k}} \freq{w}'_{km}[\nu],
\end{align}
where $\alpha_k \triangleq \sum_{m=1}^{M} \Exp \bigl[ |\freq{w}'_{km}[\nu]|^2 \bigr]$ and $\freq{w}'_{km}[\nu]$ is element $(k,m)$ of the matrix $\mathbf{\freqvec{W}}'[\nu]$.  In practice, \RZFC would always be preferred because of its superior performance.  The two other combiners, \MRC and \ZFC, are included for their mathematical tractability.  The \MRC also has an implementational advantage over the other combiners---it is possible to do most of its signal processing locally at the antennas in a distributed fashion.

\begin{remark}\label{rem:nr RxC_taps}
	As was noted in \cite{mollen2016waveforms}, the energy of the impulse response in \eqref{eq:combiner_impulse_response} is generally concentrated to a little more than $L$ of the taps for the receive combiners defined in \eqref{eq:combiners}.  For example, the energy is concentrated to exactly $L$ taps for \MRC, whose impulse response is the time-reversed impulse response of the channel.  Because, in general, $L \ll N_\text{d}$, a shorter impulse response simplifies the implementation of the receive combiner.
\end{remark}

\subsection{Quantization Error and its Effect on Single-Carrier and OFDM Transmission}\label{sec:quant_noise}
In this section, we show that the estimation error due to quantization consists of two parts: one amplitude distortion and one circularly symmetric.  The amplitude distortion degrades the performance of the \OFDM system more than it does the single-carrier system.  In a wideband system however, the amplitude distortion is negligible and \OFDM works just as well as single-carrier transmission.  

If $\{h_{mk}[\ell]\}$ is a set of uncorrelated variables, the quantization distortion can be written as:
\begin{align}\label{eq:907898191}
	e_m[n] = \sum_{k=1}^{K} \sum_{\ell=0}^{L-1} \frac{\Exp \bigl[ h^*_{mk}[\ell] e_m[n] \bigm| \{ x_k[n] \} \bigr]}{\Exp\bigl[|h_{mk}[\ell]|^2 \bigr]} h_{mk}[\ell] + d_m[n],
\end{align}
where $d_m[n]$ is the residual error with the smallest variance.  The sum in \eqref{eq:907898191} can be seen as the \LMMSE estimate of $e_m[n]$ based on $\{h_{mk}[\ell]\}$ conditioned on $x_k[n]$ and the second term as the estimation error, which is uncorrelated to the channel $\{h_{mk}[\ell]\}$.  The following lemma gives the coefficients in this sum.
\begin{lemma}\label{lem:qnoise_cond_correlation}
	If $h_{mk}[\ell] \sim \mathcal{CN}(0, p[\ell])$, the normalized conditional correlation
	\begin{align}
	\frac{\Exp \bigl[ h^*_{mk}[\ell] e_m[n] \bigm| \{ x_k[n] \} \bigr]}{\Exp\bigl[|h_{mk}[\ell]|^2 \bigr]}
	&= \sqrt{\frac{2}{\pi}} x_k[n-\ell] \tau[n] \label{eq:910982929}\\
	&\xrightarrow{\text{a.s.}}  0, \quad L\to\infty,
	\end{align}
	where
	\begin{align}
	\tau[n] \triangleq \frac{ \sqrt{P_\text{rx}[n]}}{P_\text{rx}[n]} - \frac{\Exp \Bigl[ \sqrt{P_\text{rx}[n]} \Bigr]}{\bar{P}_\text{rx}}.
	\end{align}
\end{lemma}
\begin{IEEEproof}
	See Appendix~\ref{app:qnoise_cond_correlation}.
\end{IEEEproof}

By assuming that the channel taps are uncorrelated to each other and by using \eqref{eq:910982929} in \eqref{eq:907898191}, the quantization distortion becomes:
\begin{align}\label{eq:quantization_noise_expansion}
	e_m[n] = \sqrt{\frac{2}{\pi}} \tau[n] \bar{y}_m[n] + d_m[n],
\end{align}
where the noise-free received signal is
\begin{align}
	\bar{y}_m[n] \triangleq \sum_{k = 1}^{K} \sum_{\ell = 0}^{L-1} h_{mk}[\ell] x_k[n-\ell].
\end{align}

By using \eqref{eq:quantization_noise} to write \mbox{$q_m[n] = \rho y_m[n] + e_m[n]$}, the symbol estimate of the receive combiner in \eqref{eq:RxC} can be written as:
\begin{align}\label{eq:9283490810}
	\hat{x}_k[n] = \sum_{m=1}^{M} \sum_{\ell = 0}^{N_\text{d} - 1} w_{km}[\ell] (\rho y_m[n - \ell] + e_m[n - \ell]).
\end{align}
Therfore, we define the error due to quantization as
\begin{align}
	e'_k[n] &\triangleq \sum_{m=1}^{M} \sum_{\ell = 0}^{N_\text{d} - 1} w_{km}[\ell] e_m[n - \ell]\\
	&= \sqrt{ \frac{2}{\pi}} \sum_{m=1}^{M} \sum_{\ell = 0}^{N_\text{d} - 1}  w_{km}[\ell] \tau[n - \ell] \bar{y}_{m}[n - \ell] \notag\\
	&\quad+ \sum_{m=1}^{M} \sum_{\ell = 0}^{N_\text{d} - 1} w_{km}[\ell] d_m[n {-} \ell]. \label{eq:8989023u4}
\end{align}
The first term in \eqref{eq:8989023u4} contains the noise-free received signal $\bar{y}_m[n]$ and will result in an amplitude distortion, i.e., error that contains a term that is proportional to the transmit signal $x_k[n]$ or the negative transmit signal $-x_k[n]$ (depending on the sign of $\tau[n]$).  

When the number of channel taps goes to infinity, three things happen.  (i) The amplitude distortion that contains $\tau[n]$ vanishes because $\tau[n] \to 0$ as $L \to \infty$ according to Lemma~\ref{lem:qnoise_cond_correlation}.  (ii) The variance of the error approaches
\begin{align}\label{eq:var_error_due_to_quant}
\Exp \bigl[ |e'_k[n]|^2 \bigr] \to \Exp\bigl[ |d_m[n]|^2 \bigr] = E, \quad L\to\infty.
\end{align}
(iii)  The number of terms in the second sum in \eqref{eq:8989023u4} grows with $L$, as noted in Remark~\ref{rem:nr RxC_taps}.  Therefore the sum converges in distribution to a Gaussian random variable according to the central limit theorem:
\begin{align}
	e'_k[n] \xrightarrow{\text{dist.}} \mathcal{CN}(0,  E), \quad L \to \infty.
\end{align}
The rate at which the amplitude distortion vanishes depends on the rate of convergence in \eqref{eq:817973391}, i.e., the amplitude distortion is small in systems, in which $P_\text{rx}[n]$ is close to $\bar{P}_\text{rx}$ for all $n$.

\begin{figure}
	\centering \captionsetup{position=top}
	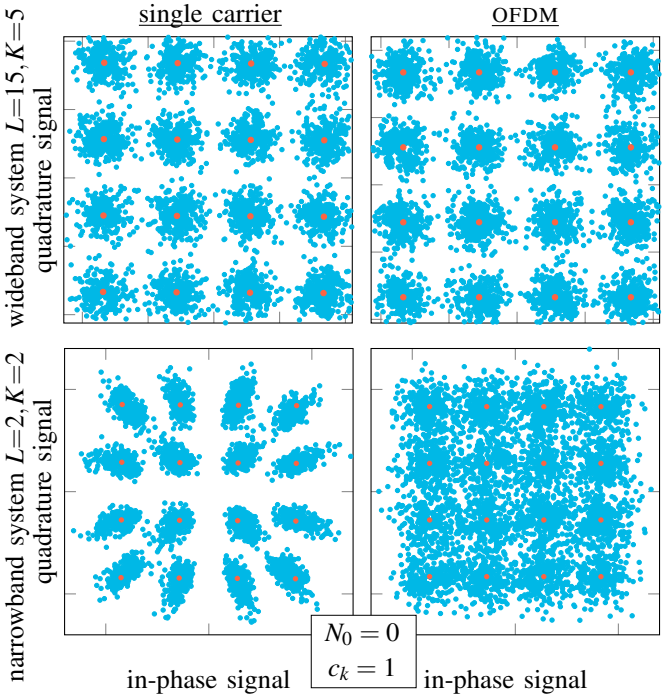
	\vspace{-0ex}
	\caption{Symbol estimates after one-bit quantization and \ZFC in a massive \MIMO base station with 128 antennas that serves $K$ users over an $L$-tap channel.  Even without thermal noise $N_0 = 0$ (the received powers $\beta_k P_k = 1$ for all users $k$) and perfect channel state information ($c_k = 1$), \ZFC cannot suppress all interference due to the quantization.  The amplitude distortion, which manifests itself as oblong clouds pointing away from the origin in the lower left narrowband system, disturbs the orthogonality of the \OFDM symbols in the lower right system and causes additional estimation error.  The amplitude distortion is negligible in the single-carrier wideband system (the quantization distortion forms circular, not oblong, clouds), which makes the estimates of the single-carrier and \OFDM systems at the top equally good.}
	\label{fig:rx_diagrams}
\end{figure}

The effect of the quantization can be seen in Figure~\ref{fig:rx_diagrams}, where the symbol estimates $\hat{s}_k[n]$ after receive combining are shown for four systems.  All other sources of estimation error (except quantization) have been removed: there is no thermal noise, no error due to imperfect channel state knowledge and \ZFC is used to suppress interuser interference.  In narrowband systems, there is coherent amplitude distortion that increases the variance of the symbol error due to quantization and that will not disappear by increasing the number of antennas.  We see that the impact of the amplitude distortion is more severe in the \OFDM system, where the it gives rise to intersymbol interference, than in the single-carrier system, where distinct, albeit non-symmetric, clusters still are visible.  In wideband systems, the amplitude distortion has vanished and there is no visible difference in the distribution of the quantization distortion of the symbol estimates for single-carrier and \OFDM transmission.  This phenomenon was studied in detail in \cite{mollen2015onebit}, where it was found that the symbol distortion due to quantization, in general, results in a nonlinear distortion of the symbol amplitudes.  If, however, the effective noise (interference plus thermal noise) is large compared to the power of the desired received signal, then the amplitude distortion vanishes and the estimated symbol constellation is a scaled and noisy version of the transmitted one.  This happens when the number of channel taps or the number of users is large.

In a wideband massive \MIMO system with one-bit \ADCs and linear combiners, there is thus no amplitude distortion and the error due to quantization can be treated as additional \AWGN (Additive White Gaussian Noise).  As a consequence, the transmission can be seen as the transmission over several parallel frequency-flat \AWGN channels.  Over such channels, the performance of different symbol constellations can be evaluated using standard methods, such as minimum Euclidean distance relative to the noise variance.  Specifically, arbitrary \QAM constellations can be used as well as \OFDM.  Detailed results with practical symbol constellations can be found in \cite{jacobsson2016massive}.
	
\subsection{Achievable Rate}	
	In this section, we derive an achievable rate for the uplink of the quantized one-bit \ADC massive \MIMO system.  The achievable rate, in the limit of a large number of channel taps $L$, is then derived in closed form.  As will be seen, this limit closely approximates the achievable rate of a wideband system also with practically large $L$.  
	
	Using the orthogonality principle, the estimate $\freq{\hat{x}}_k[\nu]$ can be written as a sum of two terms
	\begin{align}\label{eq:65432165}
		\freq{\hat{x}}_k[\nu] = a \freq{x}_k[\nu] + \freq{\zeta}_k[\nu],
	\end{align}
	where $\freq{\zeta}_k[\nu]$ is the residual error.  By choosing the factor \mbox{$a \triangleq \Exp \bigl[ \freq{x}^*_k[\nu] \freq{\hat{x}}_k[\nu] \bigr]$}, the variance of the error $\freq{\zeta}_k[\nu]$ is minimized and the error becomes uncorrelated to the transmit signal $\freq{x}_k[\nu]$.  The variance of the error term is then 
	\begin{align}
		\Exp\bigl[ |\freq{\zeta}_k[\nu]|^2 \bigr] = \Exp \bigl[ |\freq{\hat{x}}_k[\nu]|^2 \bigr] - \bigl| \Exp\bigl[ \freq{x}^*_k[\nu] \freq{\hat{x}}_k[\nu] \bigr] \bigr|^2.
	\end{align}
	
	If we denote the distribution of the transmit signal $\freq{x}_k[\nu]$ by $f_X$, an achievable rate can be derived in the following manner.  The capacity is lower bounded by:	
	\begin{align}
		C &= \max_{\{f_X : \Exp[|\freq{x}_k[\nu]|^2] \leq 1\}} I(\freq{x}_k[\nu]; \freq{\hat{x}}_k[\nu])\\
		&\geq I(\freq{x}_k[\nu]; \freq{\hat{x}}_k[\nu])\bigr\rvert_{\freq{x}_k[\nu] \sim \mathcal{CN}(0,1)}\label{eq:129871919}\\
		&\geq R_k \triangleq \log_2 \Biggl( 1 + \frac{ \bigl| \Exp \bigl[ \freq{x}^*_k[\nu] \freq{\hat{x}}_k[\nu] \bigr] \bigr|^2}{\Exp \bigl[ |\freq{\hat{x}}_k[\nu]|^2 \bigr] - \bigl| \Exp \bigl[ \freq{x}^*_k[\nu] \freq{\hat{x}}_k[\nu] \bigr] \bigr|^2 } \Biggr).\label{eq:raw_achievable_rate}
	\end{align}
	In \eqref{eq:129871919}, the capacity is bounded by assuming that the transmit signals are Gaussian.  In \eqref{eq:raw_achievable_rate}, we use results from \cite[eq.~(46)]{medard2000effect} to lower bound the mutual information.  The expectations are over the small-scale fading and over the symbols.  The derived rate is thus achievable by coding over many channel realizations.  In hardened channels \cite{1327795}, however, the rate is achievable for any single channel realization with high probability.
	
	Because the Fourier transform is unitary, the corresponding rate for single-carrier transmission is the same as \eqref{eq:raw_achievable_rate}, which can be proven by showing that $\Exp\bigl[ x^*_k[n] \hat{x}_k[n] \bigr] = \Exp\bigl[ \freq{x}^*_k[\nu] \freq{\hat{x}}_k[\nu] \bigr]$. 
	
	To gain a better understanding of the achievable rate, we will partition the estimate $\freq{\hat{x}}_k[\nu]$ into components that are uncorrelated to the transmit signal $\freq{x}_k[\nu]$.  By writing the channel as $\freq{h}_{mk}[\nu] = \freq{\hat{h}}_{mk}[\nu] + \freq{\varepsilon}_{mk}[\nu]$ the received signal becomes
	\begin{align}
		\freq{y}_m[\nu] &= \sum_{k = 1}^{K} \sqrt{c_k \beta_k P_k} \freq{\bar{y}}_{mk}[\nu] + \freq{u}_m[\nu] + \freq{z}_m[\nu],
	\intertext{where}
		\freq{\bar{y}}_{mk}[\nu] &\triangleq \frac{1}{\sqrt{c_k}} \freq{\hat{h}}_{mk}[\nu] \freq{x}_{k}[\nu],\\
		\freq{u}_m[\nu] &\triangleq \sum_{k = 1}^{K} \sqrt{\beta_k P_k} \freq{\varepsilon}_{mk}[\nu] \freq{x}_{k}[\nu].
	\end{align}
	Just like the time-domain estimate in \eqref{eq:9283490810}, the frequency-domain estimate of the transmit signal can be partitioned by rewriting the quantized signal using the relation in \eqref{eq:quantization_noise}:
	\begin{align}
		\freq{\hat{x}}_k[\nu] &= \sum_{m=1}^{M} \freq{w}_{km}[\nu] \Biggl( \rho \sum_{k'=1}^{K} \sqrt{c_{k'} \beta_{k'} P_{k'}} \freq{\bar{y}}_{mk'}[\nu] \notag\\
		&\quad+ \rho \freq{u}_m[\nu] + \rho \freq{z}_m[\nu] + \freq{e}_m[\nu]\Biggr)\\
		&= \! \rho \!\! \sum_{k' = 1}^{K}\!\!\! \sqrt{c_{k'} \! \beta_{k'} \! P_{k'} \!} \!\underbrace{\sum_{m=1}^{M} \!\! \freq{w}_{km} \! [\nu] \freq{\bar{y}}_{mk'} \! [\nu]}_{\triangleq \freq{\hat{x}}'_{kk'}[\nu]} {+} \rho \!\! \underbrace{\sum_{m=1}^{M} \!\! \freq{w}_{km} \! [\nu] \freq{u}_m \! [\nu]}_{\triangleq \freq{u}'_{k}[\nu]}\notag\\
		&\quad+ \rho \underbrace{\sum_{m=1}^{M} \freq{w}_{km}[\nu] \freq{z}_m[\nu]}_{\triangleq \freq{z}'_{k}[\nu]} + \underbrace{\sum_{m=1}^{M} \freq{w}_{km}[\nu] \freq{e}_m[\nu]}_{= \freq{e}'_{k}[\nu]}.\label{eq:many_definitions}
	\end{align}
	The terms $\{\freq{\hat{x}}'_{kk'}[\nu]\}$ can further be split up in a part that is correlated to the transmit signal and a part that is not: 
	\begin{align}\label{eq:standrad_rx_signal_splitup}
		\freq{\hat{x}}'_{kk'}[\nu] = \alpha_{kk'} \freq{x}_k[\nu] + \freq{i}_{kk'}[\nu],
	\end{align}
	where $\alpha_{kk'} \triangleq \Exp \bigl[ \freq{x}^*_k[\nu] \freq{\hat{x}}'_{kk'}[\nu] \bigr]$ and $\freq{i}_{kk'}[\nu]$ is the interference that is uncorrelated to $\freq{x}_k[\nu]$.  It is seen that $\alpha_{kk'} = 0$ for all $k' \neq k$, i.e., that only the term $\freq{\hat{x}}'_{kk}[\nu]$ is correlated to the transmit signal $\freq{x}_k[\nu]$.  We denote the gain \mbox{$G_k \triangleq |\alpha_{kk}|^2$} and the interference variance \mbox{$I_{kk'} \triangleq \Exp\bigl[|\freq{i}_{kk'}[\nu]|^2\bigr]$}.  Since they do not depend on the quality of the channel estimates nor on the quantization coarseness, they characterize the combiner that is used.  In general, these characteristic parameters are determined numerically.  Using results from random matrix theory, they were computed for \MRC and \ZFC in \cite{yang2013performance, diva2:546041} for an \IID Rayleigh fading channel $h_{mk}[\ell] \sim \mathcal{CN}(0, p[\ell])$:
	\begin{align}
	G_k = \begin{cases}
	 M\\
	 M - K
	\end{cases},\qquad
	I_{kk'} = \begin{cases}
	1, \hspace{1cm} &\text{for \MRC}\\
	0, &\text{for \ZFC}
	\end{cases}.\label{eq:characteristic_parameters_MRC_ZFC}
	\end{align}
	With \RZFC, the parameter $\lambda$ can balance array gain and interference suppression to obtain characteristic parameters in between those of \MRC and \ZFC to maximize the \SINR (Signal-to-Noise-and-Interference Ratio) of the symbol estimates that is given by the following theorem for wideband channels.
	\begin{theorem}\label{the:achievable_rate}
		When the small-scale fading coefficients are \IID and $h_{mk}[\ell] \sim \mathcal{CN}\bigl(0, p[\ell]\bigr)$, the achievable rate $R_k$ in \eqref{eq:raw_achievable_rate} approaches
		\begin{align}
			R_k \to R'_k, \quad L \to \infty,
		\end{align}
		where
		\begin{align}\label{eq:acheivable_rate}
			R'_k \triangleq \log_2 \!\left(\! 1 + \frac{c_k \beta_k P_k G_k}{\sum\limits_{\mathclap{k'=1}}^{K} \beta_{k'} P_{k'} \bigl( 1 {-} c_{k'} (1 {-} I_{kk'}) \bigr) + N_0 + Q'} \! \right).
		\end{align}
	\end{theorem}
\begin{IEEEproof}
	See Appendix~\ref{app:achievable_rate}.
\end{IEEEproof}	

	From \eqref{eq:characteristic_parameters_MRC_ZFC}, we get the following corollary about \MRC and \ZFC.
	\begin{corollary}
		The achievable rates for \MRC and \ZFC systems, where $h_{mk}[\ell] \sim \mathcal{CN}(0, p[\ell])$ \IID and when $L \to \infty$, are
		\begin{align}
			R_\text{MRC} &= \log_2 \Bigl( 1 + \frac{2}{\pi} \frac{c_k \beta_k P_k M}{N_0 + \sum_{k'=1}^{K} \beta_{k'} P_{k'}} \Bigr), \label{eq:rate_MRC}\\
			R_\text{ZFC} &= \log_2 \Bigl( 1 + \frac{2}{\pi} \frac{c_k \beta_k P_k (M - K)}{N_0  + \sum_{k'=1}^{K} \beta_{k'} P_{k'} (1 - c_{k'}\frac{2}{\pi})} \Bigr).\label{eq:rate_ZFC}
		\end{align}
	\end{corollary}

	\begin{remark}\label{rem:performance_deg_due_to_quant}
		By looking at the \SINR of \eqref{eq:acheivable_rate}, we see that, whereas the numerator scales with $G_k$, which scales with $M$ for \MRC and \ZFC, the variance of the quantization distortion $Q'$ does not scale with $M$, just like the other noise terms (which was observed in \cite{marzetta2010noncooperative} too).  In a wideband system, quantization is thus a noncoherent noise source that disappears in the limit $M \to \infty$.  Hence, arbitrary high rates are achievable by increasing the number of antennas, also in a system with one-bit \ADCs.  
	\end{remark}

	For the unquantized \MRC and \ZFC, the achievable rates become:
	\begin{align}
		R_{\text{\MRC}_0} &= \log_2 \biggl( 1 + \frac{c_k \beta_k P_k M}{N_0 + \sum_{k'=1}^{K} \beta_{k'} P_{k'}} \biggr) \label{eq:rate_MRC_unquant}\\
		R_{\text{\ZFC}_0} &= \log_2 \biggl( 1 + \frac{c_k \beta_k P_k (M - K)}{N_0 + \sum_{k'=1}^{K} \beta_{k'} P_{k'} (1 - c_{k'})} \biggr).
	\end{align}
	Note that $c_{k}$ should be understood as the channel estimation quality of the unquantized system $c_k\bigr|_{Q=0}$; it is not the same as $c_k$ in \eqref{eq:rate_MRC} and \eqref{eq:rate_ZFC}.

	\begin{remark}\label{rem:2dB_rule}
		For quantized \MRC with pilot excess factor $\mu_\text{q}$, the \SINR in \eqref{eq:rate_MRC} is a fraction
		\begin{align}\label{eq:power_loss}
			\frac{2}{\pi \Delta(\mu_0, \mu_\text{q})}
		\end{align}
		smaller than the \SINR of the unquantized system in \eqref{eq:rate_MRC_unquant} with pilot excess factor $\mu_0$ independently of the \SNR.  With equal channel estimation quality $\Delta(\mu_0, \mu_\text{q}) = 1$ the \SINR loss is $2 / \pi \approx \unit[-2]{\text{dB}}$.  In light of \eqref{eq:9982341}, the \SINR loss increases to \unit[$-$4]{dB} if both pilot excess factors $\mu_\text{q} = \mu_0 = 1$ and the receive powers $\beta_k P_k = P$ are the same from all users and the power delay profile $p[\ell] = 1/L$ for all $\ell$.  The same \SINR loss is experienced in the quantized \ZFC system at low \SNR $\beta_k P_k / N_0$.   At high \SNR however, the performance of \ZFC is greatly reduced as the interference is not perfectly suppressed.  Even with perfect channel state information ($c_k = 1$), it is seen from the rate expression \eqref{eq:rate_ZFC} that there is residual interference in the quantized system.  This gives a rate ceiling, as was pointed out in for example \cite{jindal2006mimo, ding2007multiple}.  In \cite{jindal2006mimo, ding2007multiple}, the reason for the incomplete interference suppression was imperfect channel state knowledge.  In the quantized system, the reason is the distortion of the received signals.  Whereas the rate of the unquantized \ZFC system grows without bound as $P / N_0 \to \infty$ ($\beta_k P_k = P, \forall k$), the rate of the quantized system approaches the rate ceiling:
		\begin{align}\label{eq:928923894792}
			R_\text{ZFC} \to \log_2 \Biggl( 1 + \frac{N_\text{p} (M - K)}{(\frac{\pi}{2} - 1 ) K (N_\text{p} + K)} \Biggr), \quad \frac{P}{N_0} \to \infty.
		\end{align}
		Thus, one-bit \ADCs with \ZFC work well at low \SNR, but incur a performance loss at high \SNR.  At high \SNR, however, other imperfections than quantization also limit the performance of \ZFC.  For example, pilot contamination \cite{marzetta2010noncooperative} results in a rate ceiling also in the unquantized system, which is not apparent in our analysis.  The performance loss at high \SNR might therefore be smaller than predicted here.
	\end{remark}
	
	Because of the similarities between the rate expressions of the quantized and unquantized systems, many of the properties of the unquantized massive \MIMO system carry over to the one-bit quantized system.  For example that \ZFC performs poorly when the number of antennas $M$ is close to the number of users $K$, i.e., when $M-K$ is small ($M \geq K$ for \ZFC to exist).  Similarly, quantization does not change the fact that the rate of \MRC is higher than that of \ZFC at low \SNR, where array gain, which is larger for \MRC than for \ZFC ($2M/\pi$ compared to $2(M-K)/\pi$), is more important than interference suppression.
	
	Earlier results showed that, with perfect channel state information, the capacity of a \SISO channel \cite{viterbi1979principles} and a \MIMO channel \cite{mezghani2007ultra} decreases by a factor $2 / \pi$ at low \SNR when the signals are quantized by one-bit \ADCs.  Our results indicate that the rate expressions for the low-complexity detectors \MRC and \ZFC also decrease by a factor $2 / \pi$ at low \SNR when one-bit \ADCs are used, as long as $\Delta(\mu_0, \mu_\text{q}) = 1$, i.e., as long as the channel state information is the same in the quantized and unquantized systems.  
	
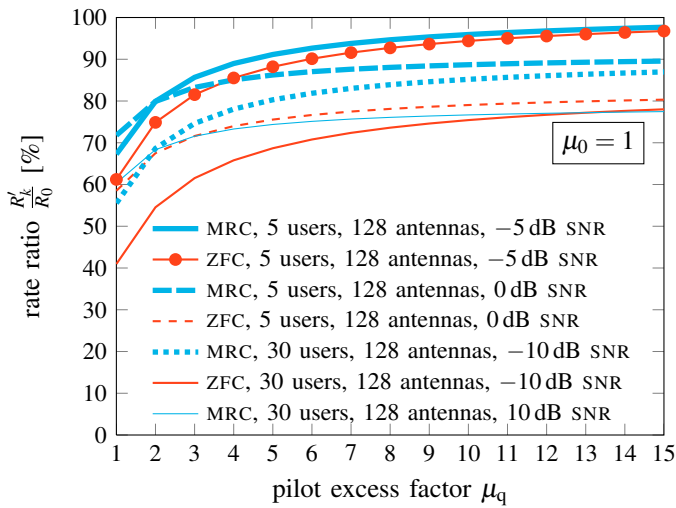
\begin{figure}
		\centering
		\scalebox{1}{\begin{tikzpicture}
		\begin{axis}[
		width=25em,
		height=45ex,
		xmin = 1,
		xmax = 15,
		xtick = {1,2,3,4,5,6,7,8,9,10,11,12,13,14,15,16,17,18,19,20},
		xticklabels = {1,2,3,4,5,6,7,8,9,10,11,12,13,14,15,16,17,18,19,20},
		ytick = {0,10,20,30,40,50,60,70,80,90,100},
		yticklabels = {0,10,20,30,40,50,60,70,80,90,100},
		ymin = 0,
		ymax = 100,
		xlabel = {pilot excess factor $\mu_\text{q}$},
		ylabel={rate ratio $\frac{R'_k}{R_0}$ [\%]},
		legend style = {cells={anchor=west}, draw=none, fill = none, anchor = south east, at = {(axis cs:14.5,0)}},
		samples at = {1,2,3,4,5,6,7,8,9,10,11,12,13,14,15,16},
		]
		\def\excessfactor{x}
		\def\power{10^(-5/10)}
		\def\nrusers{5}
		\def\nrantennae{128}
		\def\powerlossUQ{(\power * \nrusers / (\power * \nrusers + 1))}
		\def\powerlossQ{(\power * \nrusers * \excessfactor / (\power * \nrusers * \excessfactor + 1 + (pi / 2 - 1) * (1 + \nrusers * \power)))}
		\def\UQrate{(ln(1 + \powerlossUQ * \power * \nrantennae / (1 + \nrusers * \power))/ln(2))}
		\def\Qrate{(ln(1 + 2 / pi * \powerlossQ * \power * \nrantennae / (1 + \nrusers * \power))/ln(2))}
		\addplot [domain = -50:10, color = color1, line width = 2pt, samples = 100]
		{100 * \Qrate / \UQrate};
		\addlegendentry{\MRC, 5 users, 128 antennas, \unit[$-$5]{dB} \SNR};
		
		\def\nrantennae{128}
		\def\powerlossUQ{(\power * \nrusers / (\power * \nrusers + 1))}
		\def\powerlossQ{(\power * \nrusers * \excessfactor / (\power * \nrusers * \excessfactor + 1 + (pi / 2 - 1) * (1 + \nrusers * \power)))}
		\def\UQrate{(ln(1 + \powerlossUQ * \power * (\nrantennae - \nrusers) / (1 + \nrusers * \power * (1 - \powerlossUQ))) / ln(2))}
		\def\Qrate{(ln(1 + 2 / pi * \powerlossQ * \power * (\nrantennae - \nrusers) / (1 + \nrusers * \power * (1 - \powerlossQ * 2 / pi)))/ln(2))}
		\addplot [domain = -50:10, color = color2, line width = .8pt, samples = 100, mark=*] 
		{100 * \Qrate / \UQrate};
		\addlegendentry{\ZFC, 5 users, 128 antennas, \unit[$-$5]{dB} \SNR};
		
		\def\power{10^(0/10)}
		\def\nrusers{5}
		\def\nrantennae{128}
		\def\powerlossUQ{(\power * \nrusers / (\power * \nrusers + 1))}
		\def\powerlossQ{(\power * \nrusers * \excessfactor / (\power * \nrusers * \excessfactor + 1 + (pi / 2 - 1) * (1 + \nrusers * \power)))}
		\def\UQrate{(ln(1 + \powerlossUQ * \power * \nrantennae / (1 + \nrusers * \power))/ln(2))}
		\def\Qrate{(ln(1 + 2 / pi * \powerlossQ * \power * \nrantennae / (1 + \nrusers * \power))/ln(2))}
		\addplot [domain = -50:10, color = color1, dash pattern = on 0.25cm off 0.06cm, line width = 2pt, samples = 100]
		{100 * \Qrate / \UQrate};
		\addlegendentry{\MRC, 5 users, 128 antennas, \unit[0]{dB} \SNR};
		
		\def\power{10^(0/10)}
		\def\nrusers{5}
		\def\nrantennae{128}
		\def\powerlossUQ{(\power * \nrusers / (\power * \nrusers + 1))}
		\def\powerlossQ{(\power * \nrusers * \excessfactor / (\power * \nrusers * \excessfactor + 1 + (pi / 2 - 1) * (1 + \nrusers * \power)))}
		\def\UQrate{(ln(1 + \powerlossUQ * \power * (\nrantennae - \nrusers) / (1 + \nrusers * \power * (1 - \powerlossUQ)))/ln(2))}
		\def\Qrate{(ln(1 + 2 / pi * \powerlossQ * \power * (\nrantennae - \nrusers) / (1 + \nrusers * \power * (1 - \powerlossQ * 2 / pi)))/ln(2))}
		\addplot [domain = -50:10, color = color2, dashed, line width = .8pt, samples = 100]
		{100 * \Qrate / \UQrate};
		\addlegendentry{\ZFC, 5 users, 128 antennas, \unit[0]{dB} \SNR};
		
		\def\nrusers{30}
		\def\nrantennae{128}
		\def\power{10^(-10/10)}
		\def\powerlossUQ{(\power * \nrusers / (\power * \nrusers + 1))}
		\def\powerlossQ{(\power * \nrusers * \excessfactor / (\power * \nrusers * \excessfactor + 1 + (pi / 2 - 1) * (1 + \nrusers * \power)))}
		\def\UQrate{(ln(1 + \powerlossUQ * \power * \nrantennae / (1 + \nrusers * \power))/ln(2))}
		\def\Qrate{(ln(1 + 2 / pi * \powerlossQ * \power * \nrantennae / (1 + \nrusers * \power))/ln(2))}
		\addplot [domain = -50:10, color = color1, dotted, line width = 2pt]
		{100 * \Qrate / \UQrate};
		\addlegendentry{\MRC, 30 users, 128 antennas, \unit[$-$10]{dB} \SNR};
		
		\def\nrusers{30}
		\def\nrantennae{128}
		\def\powerlossUQ{(\power * \nrusers / (\power * \nrusers + 1))}
		\def\powerlossQ{(\power * \nrusers * \excessfactor / (\power * \nrusers * \excessfactor + 1 + (pi / 2 - 1) * (1 + \nrusers * \power)))}
		\def\UQrate{(ln(1 + \powerlossUQ * \power * (\nrantennae - \nrusers) / (1 + \nrusers * \power * (1 - \powerlossUQ)))/ln(2))}
		\def\Qrate{(ln(1 + 2 / pi * \powerlossQ * \power * (\nrantennae - \nrusers) / (1 + \nrusers * \power * (1 - \powerlossQ * 2 / pi)))/ln(2))}
		\addplot [domain = -50:10, color = color2, solid, line width = .8pt]
		{100 * \Qrate / \UQrate};
		\addlegendentry{\ZFC, 30 users, 128 antennas, \unit[$-$10]{dB} \SNR};
		
		\def\power{10^(10/10)}
		\def\nrusers{30}
		\def\nrantennae{128}
		\def\powerlossUQ{(\power * \nrusers / (\power * \nrusers + 1))}
		\def\powerlossQ{(\power * \nrusers * \excessfactor / (\power * \nrusers * \excessfactor + 1 + (pi / 2 - 1) * (1 + \nrusers * \power)))}
		\def\UQrate{(ln(1 + \powerlossUQ * \power * \nrantennae / (1 + \nrusers * \power))/ln(2))}
		\def\Qrate{(ln(1 + 2 / pi * \powerlossQ * \power * \nrantennae / (1 + \nrusers * \power))/ln(2))}
		\addplot [domain = -50:10, color = color1, solid, line width = .4pt]
		{100 * \Qrate / \UQrate};
		\addlegendentry{\MRC, 30 users, 128 antennas, \unit[10]{dB} \SNR};		
		
		\node[draw, anchor = north east, fill = white] at (axis cs:14.5, 75) {$\mu_0 = 1$};
		\end{axis}
		\end{tikzpicture}}
		\caption{Performance ratio $R'_k / R_0$ between the quantized and unquantized systems as a function of the pilot excess factor of the quantized system.  The pilot excess factor of the unquantized system is $\mu_0 = 1$.  All users have equal \SNR $\beta_k P_k / N_0$.  \label{fig:rate_ratio_vs_excess_factor}}
	\end{figure}

	To compare the two systems, we let $R_0$ denote the achievable rate of the unquantized system that uses a fixed pilot excess factor $\mu_0 = 1$.  The ratio $R'_k / R_0$ is drawn in Figure~\ref{fig:rate_ratio_vs_excess_factor}.  We see that the quantized system achieves approximately \unit[60--70]{\%} of the unquantized rate with \MRC in the studied systems.  With \ZFC, the ratio is around \unit[60]{\%} when there are 5~users but only \unit[40]{\%} with 30~users at low \SNR.  At high \SNR, the ratio can be much lower, e.g., \unit[20]{\%} for 30~users at \unit[10]{dB} \SNR.  Further the figure shows that the ratio can be improved by increasing the length of the pilot sequences in the quantized system.  The largest improvement, however, is by going from $\mu_\text{q} = 1$ to $\mu_\text{q} = 2$.  After that, the improvement saturates in most systems.  
	
\section{Numerical Examples}\label{sec:numerical_examples}
In this section, we verify how close the limit $R'_k$ in Theorem~\ref{the:achievable_rate} is the achievable rate $R_k$ in \eqref{eq:raw_achievable_rate} for wideband systems with a finite number of channel taps $L$.  The rate $R_k$ is numerically evaluated for the massive \MIMO system with one-bit \ADCs described in Section~\ref{sec:system_model} with the linear channel estimation and receive combining described in the Sections~\ref{sec:channel_estimation} and \ref{sec:uplink_transmission}.  As a way of comparing the quantized system to the unquantized, the number of extra antennas needed to make the quantized rate the same as the unquantized, while maintaining the same transmit power, is established.  Such a comparison is sensible in a system where the number of users is fixed.  If more users were available, a system with more antennas could potentially also serve more users and thus get a higher multiplexing gain.

The channel taps are modeled as \IID Rayleigh fading and follow a uniform power delay profile, i.e., $h_{mk}[\ell] \sim \mathcal{CN}(0, 1 / L)$.  The large-scale fading is neglected and all received powers $\beta_k P_k / N_0$ are assumed to be equal for all users $k$ in the first part of the study.  This corresponds to doing a fair power control among the users, where the transmit power $P_k$ is chosen proportional to $1 / \beta_k$.  Such a power control is possible to do since the users are assumed to know the large-scale fading.  It is also desirable many times to ensure that all served users have similar \SNR so that quality of service is uniformly good.
	
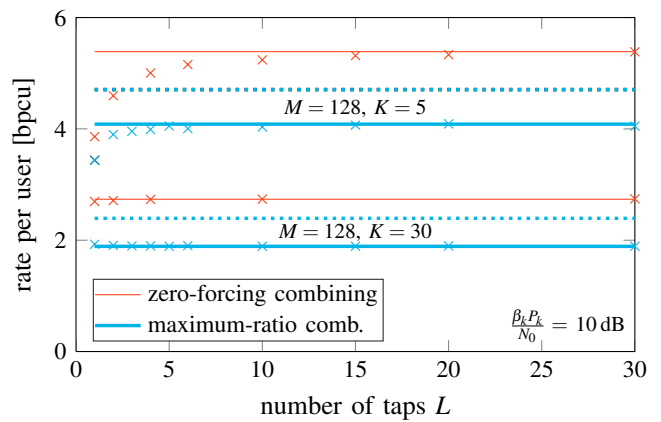
\begin{figure}
		\centering
		\setlength{\figurewidth}{22em}
		\setlength{\figureheight}{28ex}
		\scalebox{1}{

\definecolor{mycolor1}{rgb}{0.00000,0.72266,0.90234}%
\definecolor{mycolor2}{rgb}{1.00000,0.39216,0.25882}%
\begin{tikzpicture}

\begin{axis}[%
width=0.951\figurewidth,
height=\figureheight,
at={(0\figurewidth,0\figureheight)},
scale only axis,
xmin=0,
xmax=30,
xlabel={number of taps $L$},
ymin=0,
ymax=6,
ytick={0,2,4,6},
ylabel={rate per user [bpcu]},
axis background/.style={fill=white},
legend style = {at = {(0.03,0.03)}, anchor = south west, legend cell align = left, align = left, draw = white!15!black, inner sep = 0pt}
]
\addplot[only marks, mark = x, mark size = 2.4pt, color=black, forget plot]table[row sep=crcr]{1	3.43680752053469\\};\label{plo:markerx2}
\addplot [color = black, solid, forget plot]
table[row sep=crcr]{%
	1	1.88973119765006\\
	2	1.88973119765006\\
};\label{plo:solid_line}

\addplot [color=black,line width = 1.3pt,dotted, forget plot]
table[row sep=crcr]{%
	1 4.7059\\
	30 4.7059\\
};\label{plo:dotted_line}


\addplot [color=color2,solid]
table[row sep=crcr]{%
	1	5.387\\
	30	5.387\\
};
\addlegendentry{zero-forcing combining};

\addplot [only marks, mark = x, mark size = 2.4pt,color=color2,solid, forget plot]
  table[row sep=crcr]{%
  	  	1	3.86047212070494\\
  	  	2	4.59575340195669\\
  	  	4	5.0030\\
  	  	6	5.15582544234862\\
  	  	10	5.2396\\
  	  	15	5.31548034185363\\
  	  	20	5.3338\\
  	  	30	5.3844\\
};

\addplot [color=color2,line width = .8pt, dotted, forget plot]
table[row sep=crcr]{%
	1 10.266\\
	30 10.266\\
};

\addplot [color = color1, solid, line width = 1.3pt]
table[row sep=crcr]{%
	1	4.0847694232621\\
	30	4.0847694232621\\
};
\addlegendentry{maximum-ratio comb.};

\addplot [only marks, mark = x, mark size = 2.4pt,color=color1,solid, forget plot]
table[row sep=crcr]{%
	1	3.43680752053469\\
	2	3.89887133889207\\
	3	3.95621501113106\\
	4	3.98869806173608\\
	5	4.05190271806189\\
	6	4.00476691660377\\
	10	4.03447430538174\\
	15	4.06767737454789\\
	20	4.08993657212615\\
	30	4.05018785118177\\
};

\addplot [color=color1,line width = 1.3pt,dotted]
table[row sep=crcr]{%
	1 4.7059\\
	30 4.7059\\
};

\addplot [color=color1,solid,forget plot, line width = 1.3pt]
table[row sep=crcr]{%
1	1.88973119765006\\
30	1.88973119765006\\
};

\addplot [only marks, mark = x, mark size = 2.4pt,color = color1, solid, forget plot]
table[row sep=crcr]{%
	1	1.92656207896423\\
	2	1.90624438165947\\
	3	1.89850364934257\\
	4	1.89856519000916\\
	5	1.89150722498243\\
	6	1.90170937192768\\
	10	1.89197837431031\\
	15	1.89286438585175\\
	20	1.89743465073129\\
	30	1.89119851766236\\
};

\addplot [color=color2,solid,forget plot]
table[row sep=crcr]{%
	1	2.73596717013883\\
	30	2.73596717013883\\
};

\addplot [only marks, mark = x, mark size = 2.4pt, color = color2, solid, forget plot]
table[row sep=crcr]{%
	1	2.695283904517\\
	2	2.71202130314872\\
	4	2.73273966648692\\
	10	2.73700994082739\\
	30	2.74217644819692\\
};

\addplot [color=mycolor1,line width = 1.3pt, dotted]
table[row sep=crcr]{%
1 2.393\\
30 2.393\\
};

\addplot [color=mycolor2,line width = .8pt, dotted]
table[row sep=crcr]{%
1 9.9381\\
30 9.9381\\
};

\node at (axis cs:15,4.37){\footnotesize $M = 128$, $K=5$}; 
\node at (axis cs:15,2.14){\footnotesize $M = 128$, $K = 30$};

\node[font=\footnotesize, anchor=south east] at (current axis.south east) {$\frac{\beta_k P_k}{N_0} = $ \unit[10]{dB}};

\end{axis}

\end{tikzpicture}%
}
		\caption{The achievable rate $R_k$ marked \ref{plo:markerx2} and its limit $R'_k$ drawn with a solid line for a system with 128 antennas that serves 5 and 30 users over an $L$-tap channel with Rayleigh fading taps.  The dotted line shows the rate of the unquantized system with \MRC.  The rate of the unquantized \ZFC is \unit[10]{bpcu} for $K = 5$ and \unit[9.9]{bpcu} for $K = 30$.  The channel is known perfectly by the base station.}
		\label{fig:rate_vs_L}
	\end{figure}

First, we study the convergence of the achievable rate $R_k$ in \eqref{eq:raw_achievable_rate} towards its limit by comparing $R_k$ for finite $L$ to the limit $R'_k$ in \eqref{eq:acheivable_rate} in Figure~\ref{fig:rate_vs_L}.  It is seen that the limit $R'_k$ is indeed an accurate approximation of the achievable rate $R_k$ when the number of channel taps is large.  For the system with 128 antennas and 5 users, the limit $R'_k$ is close to $R_k$ already at $L = 15$~taps, which corresponds to a moderately frequency-selective channel.  For the system with 128 antennas and 30 users, however, the limit $R'_k$ is a good approximation for $R_k$ also in a narrowband scenario with $L = 1$.  This immediate convergence was explained by Remark~\ref{rem:big_K}, where it was noted that the wideband approximation is valid also when the number of users is large and there is no dominant user.

	The lower performance for small $L$ for the case of 5~users in Figure~\ref{fig:rate_vs_L} is caused by the amplitude distortion discussed in Section~\ref{sec:quant_noise}.  As the amplitude distortion disappears with more taps, the rate $R_k$ increases.  The improvement saturates when the amplitude distortion is negligible and the limit $ R'_k$ is a close approximation of $R_k$.  This suggests that linear receivers for one-bit \ADCs work better with frequency-selective channels than with frequency-flat channels and that wideband systems are beneficial when one-bit \ADCs are used.  
	
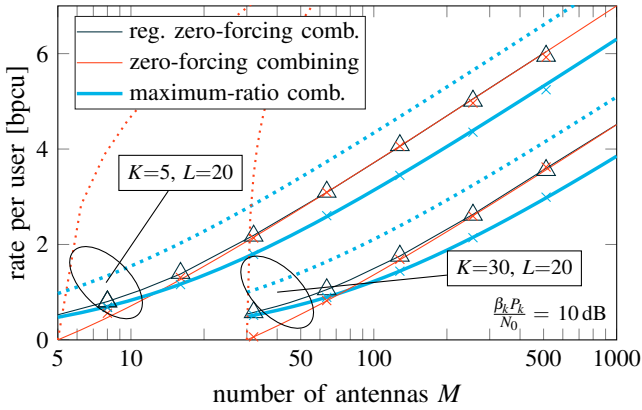
\begin{figure}
		\centering
		\setlength{\figurewidth}{22em}
		\setlength{\figureheight}{28ex}
		\scalebox{1}{

\definecolor{mycolor1}{rgb}{0.00000,0.72266,0.90234}%
\definecolor{mycolor2}{rgb}{1.00000,0.39216,0.25882}%
\begin{tikzpicture}

\begin{semilogxaxis}[%
width=0.951\figurewidth,
height=\figureheight,
at={(0\figurewidth,0\figureheight)},
scale only axis,
xmin=5,
xmax=1000,
xlabel={number of antennas $M$},
xtick = {5,6,7,8,9,10, 20,30,40,50,60,70,80,90,100,200,300,400,500, 600,700,800,900,1000},
xticklabels={5,,,,,10,,,,50,,,,,100,,,,500,,,,,1000},
ytick={0,2,4,6,8,10},
ymin=0,
ymax=7,
ylabel={rate per user [bpcu]},
ylabel style={yshift=-1ex},
axis background/.style={fill=white},
legend pos=north west,
legend style={legend cell align=left,align=left,draw=white!15!black, inner sep = 0pt},
]
\addplot[only marks, mark = x, mark size = 2.4pt, color = black, forget plot]table[row sep=crcr]{8	   0.6831\\};\label{plo:markerx3}
\addplot[only marks, mark = triangle, mark size = 4pt, color = black, forget plot]table[row sep=crcr]{8    0.8042\\};\label{plo:markertriangle}

\def\powerdB{10}
\def\power{10 ^ (\powerdB / 10)}

\addplot [only marks, mark = triangle, mark size = 4pt, color = color3, solid, forget plot]
table[row sep=crcr]{%
	8    0.8042\\
	16    1.4015\\
	32   2.1820\\
	64    3.1031\\
	128    4.0790\\
	256    5.0101\\
	512    5.9514\\
	1024    6.4631\\
};

\addplot [color = color3, solid]
table[row sep=crcr]{%
	5	0.5259\\
	8    0.8035\\
	11	1.0453\\
	16    1.3816\\
	32   2.1911\\
	64    3.0909\\
	128    4.0548\\
	256    5.0518\\
	512    6.0405\\
};
\addlegendentry{reg.\ zero-forcing comb.};

\addplot [only marks, mark = x, mark size = 2.4pt, color = color2, solid, forget plot]
  table[row sep=crcr]{%
8    0.5587\\
16    1.3080\\
32   2.1349\\
64    3.0933\\
128    4.0596\\
256    4.9639\\
512    5.9242\\
1024    -inf\\
};

\def\nrusers{5}
\def\powerloss{2 / pi * \power * \nrusers / (\nrusers * \power + 1)}
\addplot [domain=\nrusers:1000, color = color2, solid]
{ln(1 + (2 / pi) * (x - \nrusers) * \powerloss * \power / (1 + \nrusers * \power * (1 - \powerloss * 2 / pi)) ) / ln(2)};
\addlegendentry{zero-forcing combining};

\def\nrusers{5}
\def\powerloss{\power * \nrusers / (\nrusers * \power + 1)}
\addplot [domain=\nrusers:1000, color = color2, line width = .8pt, dotted, forget plot]
{ln(1 + (x - \nrusers) * \powerloss * \power / (1 + \nrusers * \power * (1 - \powerloss)) ) / ln(2)};

\addplot [only marks, mark = x, mark size = 2.4pt, color = color1, solid, forget plot]
table[row sep=crcr]{%
	8	   0.6831\\
	16	   1.1590\\
	32	   1.7904\\
	64	   2.6032\\
	128	   3.4467\\
	256	   4.3532\\
	512	   5.2424\\
	2024	   6.5016\\
};

\def\nrusers{5}
\def\powerloss{2 / pi * \power * \nrusers / (\nrusers * \power + 1)}
\addplot [domain=\nrusers:1000, color = color1, solid, line width = 1.3pt]
{ln(1 + (2 / pi) * x * \powerloss * \power / (1 + \nrusers * \power) ) / ln(2)};
\addlegendentry{maximum-ratio comb.};

\def\nrusers{5}
\def\powerloss{\power * \nrusers / (\nrusers * \power + 1)}
\addplot [domain=\nrusers:1000, color = color1, line width = 1.3pt, dotted, forget plot]
{ln(1 + x * \powerloss * \power / (1 + \nrusers * \power) ) / ln(2)};

\addplot [only marks, mark = triangle, mark size = 4pt, color = color3, solid, forget plot]
table[row sep=crcr]{%
	32   0.5780\\
	64    1.0585\\
	128    1.7616\\
	256    2.6204\\
	512    3.5572\\
	1024    4.5572\\
};

\addplot [color = color3, solid, forget plot]
table[row sep=crcr]{%
	30	 0.5426\\
	32   0.5757\\
	40	0.7055\\
	45	0.7812\\
	55	0.9299\\
	64    1.0543\\
	70	1.1324\\
	80	1.2601\\
	128    1.7567\\
	256    2.6302\\
	512    3.5961\\
	1024    4.5477\\
};

\addplot [only marks, mark = x, mark size = 2.4pt, color = color1, solid, forget plot]
table[row sep=crcr]{%
32    0.5159\\
64    0.8918\\
128    1.4400\\
256    2.1423\\
512    2.9915\\
1024    3.8985\\
};

\def\nrusers{30}
\def\powerloss{2 / pi * \power * \nrusers / (\nrusers * \power + 1)}
\addplot [domain = \nrusers:1000, color = color1, solid, forget plot, line width = 1.3pt]
{ln(1 + (2 / pi) * x * \powerloss * \power / (1 + \nrusers * \power) ) / ln(2)};

\def\nrusers{30}
\def\powerloss{\power * \nrusers / (\nrusers * \power + 1)}
\addplot [domain = \nrusers:1000, color = color1, line width = 1.3pt, dotted, forget plot]
{ln(1 + x * \powerloss * \power / (1 + \nrusers * \power) ) / ln(2)};

\addplot [only marks, mark = x, mark size = 2.4pt, color = color2, solid, forget plot]
table[row sep=crcr]{%
32    0.0631\\
64    0.8236\\
128    1.6890\\
256    2.5943\\
512    3.6237\\
1024    4.5235\\
};

\def\nrusers{30}
\def\powerloss{2 / pi * \power * \nrusers / (\nrusers * \power + 1)}
\addplot [domain = \nrusers:1000, color = color2, solid, forget plot]
{ln(1 + (2 / pi) * (x - \nrusers) * \powerloss * \power / (1 + \nrusers * \power * (1 - \powerloss * 2 / pi)) ) / ln(2)};

\def\nrusers{30}
\def\powerloss{\power * \nrusers / (\nrusers * \power + 1)}
\addplot [domain = \nrusers:1000, color = color2, line width = .8pt, dotted, samples at = {30,31,32,33,34,35,36,37,38,39,40,41,42,43,45,50,60,70}]
{ln(1 + (x - \nrusers) * \powerloss * \power / (1 + \nrusers * \power * (1 - \powerloss)) ) / ln(2)};

\node[coordinate] (A) at (axis cs:7.9,1.2){}; 
\node[draw] (Atext) at (axis cs:16,3.5){\footnotesize $K{=}5$, $L{=}20$}; 
\draw (A) -- (Atext);

\node[coordinate] (B) at (axis cs:40,1){}; 
\node[draw, right] (Btext) at (axis cs:200,1.5){\footnotesize $K{=}30$, $L{=}20$}; 
\draw (B) -- (Btext);

\node[font=\footnotesize, anchor=south east] at (current axis.south east) {$\frac{\beta_k P_k}{N_0} = $ \unit[\powerdB]{dB}};
\end{semilogxaxis}
\draw[rotate=-45] (A) ellipse (1.7em and 0.9em); 
\draw[rotate=-45] (B) ellipse (1.7em and 0.9em); 
\end{tikzpicture}%
}
		\caption{The achievable rate $R_k$ (marked \ref{plo:markerx2} and \ref{plo:markertriangle}), its limit $R'_k$ (solid lines) and the rate $R_k$ for the unquantized system (dotted lines) at high \SNR $\beta_k P_k / N_0 = $ \unit[10]{dB}, $\forall k$, using the same number of pilot symbols.  The channel taps are \IID Rayleigh fading and estimated with $N_\text{p} = KL$ pilot symbols.  The curves for single-carrier and \OFDM transmission coincide both for maximum-ratio and zero-forcing combining. }
		\label{fig:rate_vs_M}
	\end{figure}

\begin{figure}
			\setlength{\figurewidth}{22em}
			\setlength{\figureheight}{28ex}
		\centering
		\scalebox{1}{

\definecolor{mycolor1}{rgb}{0.00000,0.72266,0.90234}%
\definecolor{mycolor2}{rgb}{1.00000,0.39216,0.25882}%
\definecolor{LiUgrey}{HTML}{687F91}
\begin{tikzpicture}

\begin{semilogxaxis}[%
width=0.951\figurewidth,
height=\figureheight,
at={(0\figurewidth,0\figureheight)},
scale only axis,
xmin=5,
xmax=1000,
xlabel={number of antennas $M$},
xtick = {5,6,7,8,9,10, 20,30,40,50,60,70,80,90,100,200,300,400,500, 600,700,800,900,1000},
xticklabels={5,,,,,10,,,,50,,,,,100,,,,500,,,,,1000},
ytick={0,1,2,3,4},
ymin=0,
ymax=3.5,
ylabel={rate per user [bpcu]},
ylabel style={yshift=-1ex},
axis background/.style={fill=white},
legend pos=north west,
legend style={legend cell align=left,align=left,draw=white!15!black, inner sep = 0pt}
]
\addplot[only marks, mark = x, mark size = 2.4pt, color=black, forget plot]table[row sep=crcr]{8 0.1008\\};\label{plo:markerx}
\addplot [only marks, mark = triangle, mark size = 4pt, color = color3, solid, forget plot]
table[row sep=crcr]{%
	8  	0.1085\\
	16  	0.1657\\
	32    0.3511\\
	64    0.6658\\
	128    1.1269\\
	256    1.7597\\
	512    2.5139\\
	1024    3.5066\\
};

\addplot [color = color3, solid]
table[row sep=crcr]{%
8    0.0978\\
16    0.1892\\
32    0.3611\\
64   0.6554\\
80    0.7883\\
128    1.1214\\
180   1.4097\\
200    1.5128\\
256    1.7524\\
400    2.2432\\
512    2.5388\\
1024    3.4137\\
};
\addlegendentry{reg.\ zero-forcing comb.};

\addplot [only marks, mark = x, mark size = 2.4pt, color = color1, solid, forget plot]
table[row sep=crcr]{%
8    0.1008\\
16    0.2006\\
32    0.3552\\
64    0.6687\\
128    1.0757\\
256    1.7335\\
512    2.4556\\
1024    3.3470\\
};

\def\nrusers{5}
\def\power{10 ^ (-10 / 10)}
\def\powerloss{2 / pi * \power * \nrusers / (\nrusers * \power + 1)}
\addplot [domain = \nrusers:1000, color = color1, solid, line width = 1.3pt]
{ln(1 + (2 / pi) * x * \powerloss * \power / (1 + \nrusers * \power) ) / ln(2)};
\addlegendentry{maximum-ratio comb.};

\def\nrusers{5}
\def\power{10 ^ (-10 / 10)}
\def\powerloss{\power * \nrusers / (\nrusers * \power + 1)}
\addplot [domain = \nrusers:1000, color = color1, line width = 1.3pt, dotted, forget plot]
{ln(1 + x * \powerloss * \power / (1 + \nrusers * \power) ) / ln(2)};

\addplot [only marks, mark = x, mark size = 2.4pt, color = color2, solid, forget plot]
  table[row sep=crcr]{%
8  	0.0417\\
16  	0.1426\\
32    0.3293\\
64    0.6451\\
128    1.1294\\
256    1.7221\\
512    2.5028\\
1024    3.3903\\
};

\def\nrusers{5}
\def\power{10 ^ (-10 / 10)}
\def\powerloss{2 / pi * \power * \nrusers / (\nrusers * \power + 1)}
\addplot [domain = \nrusers:1000, color = color2, solid]
{ln(1 + (2 / pi) * (x - \nrusers) * \powerloss * \power / (1 + \nrusers * \power * (1 - \powerloss * 2 / pi)) ) / ln(2)};
\addlegendentry{zero-forcing combining};

\def\nrusers{5}
\def\power{10 ^ (-10 / 10)}
\def\powerloss{\power * \nrusers / (\nrusers * \power + 1)}
\addplot [domain = \nrusers:1000, color = color2, line width = .8pt, dotted, forget plot]
{ln(1 + (x - \nrusers) * \powerloss * \power / (1 + \nrusers * \power * (1 - \powerloss)) ) / ln(2)};


\addplot [only marks, mark = triangle, mark size = 4pt, color = color3, solid, forget plot]
table[row sep=crcr]{%
32	0.0832\\
64	0.1702\\
128	0.3315\\
256	0.5451\\
512	0.9884\\
};

\addplot [color = color3, solid, forget plot]
table[row sep=crcr]{%
5    0.0104\\
16    0.0356\\
32    0.0739\\
64    0.1542\\
128    0.3059\\
200	0.4640\\
256   0.5718\\
350	0.7424\\
400	0.8230\\
512    0.9844\\
650	1.1705\\
1024    1.5355\\
};

\addplot [only marks, mark = x, mark size = 2.4pt, color = color1, solid, forget plot]
table[row sep=crcr]{%
8 0.0051\\
16 0.0378\\
32 0.0413\\
64 0.1342\\
128 0.2378\\
256 0.4438\\
512 0.7805\\
1024 1.1485\\
};

\def\nrusers{5}
\def\powerweak{10 ^ (-10 / 10)}
\def\power{(10 * \powerweak)}
\def\powerloss{(\powerweak * \nrusers / (\powerweak * \nrusers + 1 + (pi/2 - 1 + (\nrusers - 1) * \power * (pi / 2 - 1) + \powerweak * (pi / 2 - 1))))}
\addplot [domain = \nrusers:1000, color = color1, solid, forget plot, line width = 1.3pt]
{ln(1 + (2 / pi) * x * \powerloss * \powerweak / (1 + (\nrusers - 1) * \power + \powerweak) ) / ln(2)};

\def\nrusers{5}
\def\powerloss{\power * \nrusers / (\nrusers * \power + 1)}
\addplot [domain = \nrusers:1000, color = color1, line width = 1.3pt, dashed, forget plot]
{ln(1 + x * \powerloss * \powerweak / (1 + (\nrusers - 1) * \power + \powerweak) ) / ln(2)};

\addplot [only marks, mark = x, mark size = 2.4pt, color = color2, solid, forget plot]
table[row sep=crcr]{%
8 0.0119\\
16 0.0197\\
32 0.0859\\
64 0.1715\\
128 0.2718\\
256 0.5756\\
512 0.9676\\
1024 1.6372\\
};

\def\nrusers{5}
\def\powerweak{10 ^ (-10 / 10)}
\def\power{10 * \powerweak}
\def\powerlossweak{\powerweak * \nrusers / (\powerweak * \nrusers + 1 + (1 + (\nrusers - 1) * \power * (pi / 2 - 1) + \powerweak * (pi / 2 - 1)))}
\def\powerloss{\power * \nrusers / (\power * \nrusers + 1 + (1 + (\nrusers - 1) * \power * (pi / 2 - 1) + \powerweak * (pi / 2 - 1)))}
\addplot [domain = \nrusers:1000, color = color2, solid, forget plot]
{ln(1 + (2 / pi) * (x - \nrusers) * \powerlossweak * \powerweak / (1 + (\nrusers - 1) * \power * (1 - 2 / pi * \powerloss) + \powerweak * (1 - \powerlossweak * 2 / pi)) ) / ln(2)};

\def\nrusers{5}
\def\powerlossweak{\powerweak * \nrusers / (\nrusers * \powerweak + 1)}
\def\powerloss{\power * \nrusers / (\nrusers * \power + 1)}
\addplot [domain=\nrusers:1000, color = color2, line width = .6pt, dashed, forget plot, dash pattern = on 7pt off 3pt]
{ln(1 + (x - \nrusers) * \powerlossweak * \powerweak / (1 + (\nrusers - 1) * \power * (1 - \powerloss) + \powerweak * (1 - \powerlossweak)) ) / ln(2)};

\node[left, inner sep = 0pt] (B) at (axis cs:50,1.3) {\footnotesize $K{=}5$, $L{=}20$, equal \SNR};
\draw[line width = 0.2pt] (axis cs:90,0.82) -- (B.south east);

\node[left, inner sep = 0pt] (C) at (axis cs:80,2) {\footnotesize $K{=}5$, $L{=}20$, weak user};
\draw[line width = 0.2pt] (axis cs: 195,0.38) -- (C.south east);

\node[font=\footnotesize, anchor=south east] at (current axis.south east) {$\frac{\beta_1 P_1}{N_0} = $ \unit[$-$10]{dB}};

\end{semilogxaxis}

\end{tikzpicture}%
}
		\caption{Same setup as in Figure~\ref{fig:rate_vs_M} except the \SNR is low.  For one set of curves, all users have the same \SNR $\beta_k P_k / N_0 = $ \unit[$-$10]{dB}.  For the other, marked “weak user”, the studied user has \unit[$-$10]{dB} \SNR while the interfering users have \unit[0]{dB} \SNR.  The rate of the unquantized system is drawn with dotted lines for equal \SNR and with dashed lines for the weak user.\label{fig:low_SNR}}
	\end{figure}
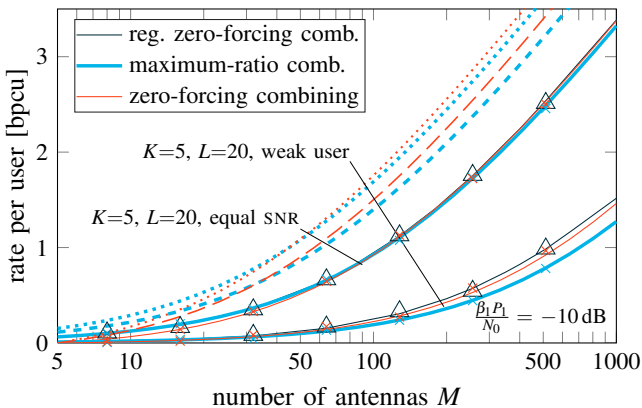

	The rates of some wideband massive \MIMO systems at high \SNR $\beta_k P_k / N_0 = \unit[\text{10}]{\text{dB}}$ and low \SNR \unit[$-$10]{dB} are shown for different numbers of base station antennas in Figures~\ref{fig:rate_vs_M} and \ref{fig:low_SNR} respectively.  We observe that the limit $R'_k$ approximates the rate $R_k$ well in all studied cases.  Furthermore, we note that the quantized system needs 2.5 times (\unit[$\approx$4]{dB}) more antennas to ensure the same rate as the unquantized system with \MRC, which was predicted in Remark~\ref{rem:2dB_rule}.  With \ZFC at high \SNR, the gap between the quantized and unquantized rates is much larger.  At low \SNR however, the gap is greatly decreased; then 2.6 times more antennas are needed in the quantized system to obtain the same performance as the unquantized system.  The rate of \RZFC is similar to \MRC and \ZFC, whichever is better for a given $M$; it is in part or fully hidden by the curves of \MRC and \ZFC.  For this reason, only the quantized \RZFC is included.
	
	In Figure~\ref{fig:low_SNR}, we consider a user whose \SNR is \unit[10]{dB} weaker than the \SNR{}s of the interfering users $\beta_k P_k / N_0 = 10 \beta_1 P_1 / N_0 = $ \unit[0]{dB} for $k = 2, 3, 4, 5$.  This can happen if there is one user whose transmit power is limited for some practical reason or if a user happens to experience shadowing by the environment.  The result is marked with “weak user” in Figure~\ref{fig:rate_vs_M}.  We see that such a weak user gets a much lower rate than the case where all users have the same \SNR.  This is because of the increased interference that the weak user suffers.  The gap between the unquantized system and the quantized system is larger for a weak user than for users with the same \SNR as all the other users because the channel estimation quality is heavily degraded when the orthogonality of the pilots is lost in the quantization.  For \ZFC, 10.4 times more antennas are needed and for \MRC 10.6 times, which should be compared to 2.6 and 2.5 times for equal \SNR.  Users that are relatively weak compared to interfering users should therefore be avoided in one-bit \ADC systems, for example by proper user scheduling.  In case weak users cannot be avoided, such users will have to obtain good channel estimates, either by longer pilot sequences or by increasing the transmit power of their pilots.

	\begin{figure}
		\centering
		\scalebox{1}{\begin{tikzpicture}
		\begin{axis}[
		width=25em,
		height=40ex,
		xmin = -20,
		xmax = 10,
		ymin = 0,
		ymax = 6,
		xlabel = {\SNR $\frac{\beta_k P_k}{N_0}$ [dB]},
		ylabel={rate per user [bpcu]},
		legend style = {cells={anchor=west}, draw=none, anchor=north west, at={($(rel axis cs:0,1) + (0.1mm, -1mm)$)}},
		]
		\def\nrusers{5}
		\def\nrantennae{128}
		\def\powerlossUQ{(10^(x/10) * \nrusers / (10^(x/10) * \nrusers + 1))}
		\def\powerlossQ{(10^(x/10) * \nrusers / (10^(x/10) * \nrusers + 1 + (pi / 2 - 1) * (1 + \nrusers * 10^(x/10))))}
		\addplot [domain = -50:10, color = color2, samples = 100]
		{ln(1 + 2 / pi * \powerlossQ * 10^(x/10) * (\nrantennae - \nrusers) / (1 + \nrusers * 10^(x/10) * (1 - \powerlossQ * 2 / pi)))/ln(2)};
		\addlegendentry{quant. \ZFC, \nrusers{} users, \nrantennae{} antennas};
		
		\addplot [domain = -50:10, color = color2, line width = .8pt, dotted]
		{ln(1 + \powerlossUQ * 10^(x/10) * (\nrantennae - \nrusers) / (1 + \nrusers * 10^(x/10) * (1 - \powerlossUQ)))/ln(2)};
		\addlegendentry{unquant.\ \ZFC, \nrusers{} users, \nrantennae{} antennas};		
		
		\addplot [domain = -50:10, color = color1, line width = 1.3pt, samples = 100]
		{ln(1 + 2 / pi * \powerlossQ * 10^(x/10) * \nrantennae / (1 + \nrusers * 10^(x/10)))/ln(2)};
		\addlegendentry{quant.\ \MRC, \nrusers{} users, \nrantennae{} antennas};
		
		\addplot [only marks, mark = x, mark size = 2.4pt,color=color1,solid, forget plot]
		table[row sep=crcr]{%
			-15	0.2604\\
			-10	1.1045\\
			-5	2.2856\\
			0	3.0387\\
			7	3.3889\\
		};
		
		\addplot [only marks, mark = x, mark size = 2.4pt,color=color2,solid, forget plot]
		table[row sep=crcr]{%
			-15	0.2475\\
			-10	1.1172\\
			-5	2.4162\\
			0	3.3891\\
			7	3.9443\\
		};		
		
		\addplot [domain = -50:10, color = color1, line width = 1.3pt, dotted, samples = 100]
		{ln(1 + \powerlossUQ * 10^(x/10) * \nrantennae / (1 + \nrusers * 10^(x/10)))/ln(2)};
		\addlegendentry{unquant.\ \MRC, \nrusers{} users, \nrantennae{} antennas};
		
		\end{axis}
		\end{tikzpicture}}
		\vspace{-0ex}
		\caption{The rate $R'_k$.  All users have the same \SNR.  The channel is estimated with $N_\text{p} = KL$ pilot symbols. \label{fig:rate_vs_power}}
	\end{figure}
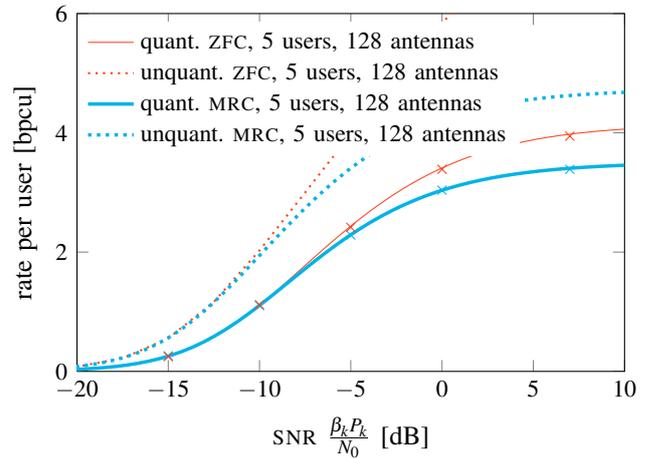	
	
	In Figure~\ref{fig:rate_vs_power}, the rate as a function of \SNR is shown for some systems.  It can be seen that the rate of the quantized systems is limited by a rate ceiling, as was indicated in \eqref{eq:928923894792}.  Around  \unit[70]{\%} of the performance of the unquantized system can be achieved by the quantized system with \MRC at \unit[$-$5]{dB} \SNR, which gives approximately \unit[2]{bpcu}.  At the same \SNR, \ZFC achieves \unit[60]{\%} of the unquantized rate, which agrees with Figure~\ref{fig:rate_ratio_vs_excess_factor}.  As observed, this performance loss can be compensated for by increasing the number of base station antennas.  An increase of antennas, however, would also lead to an increase in hardware complexity, cost and power consumption.  Having in mind that one-bit \ADCs at the same time greatly reduces these three practical issues, it is difficult to give a straightforward answer to whether one-bit \ADCs are better, in some sense, than \ADCs of some other resolution.  A thorough future study of the receive chain hardware has to answer this question.	
	
	\section{Conclusion}\label{sec:conclusion}
	We derived an achievable rate for a practical linear massive \MIMO system with one-bit \ADCs with estimated channel state information and a frequency-selective channel with \IID Rayleigh fading taps and a general power delay profile. The derived rate is a lower bound on the capacity of a massive \MIMO system with one-bit \ADCs.  As such, other nonlinear detection methods could perform better at the possible cost of increased computational complexity.  The rate converges to a closed-form limit as the number of taps grows.  We have shown in numerical examples that the limit approximates the achievable rate well also for moderately frequency-selective channels with finite numbers of taps.
	
	A main conclusion is that frequency-selective channels are beneficial when one-bit \ADCs are used at the base station.  Such channels spread the received interference evenly over time, which makes the estimation error due to quantization additive and circularly symmetric.  This makes it possible to use low-complexity receive combiners and low-complexity channel estimation for multiuser symbol detection.  One-bit \ADCs decrease the power consumption of the analog-to-digital conversion at a cost of an increased required number of antennas or reduced rate performance.  At low to moderate \SNR, approximately three times more antennas are needed at the base station to reach the same performance as an unquantized system when the channel is estimated by the proposed low-complexity channel estimation method.  
	
	The symbol estimation error due to quantization consists of two parts in a massive \MIMO system: one amplitude distortion and one additive circularly symmetric Gaussian distortion.  The amplitude distortion becomes negligible in a wideband system, which makes the implementation of \OFDM straightforward.  Since the error due to quantization is circularly symmetric Gaussian, systems that use \OFDM are affected in the same way by one-bit quantizers as single-carrier systems, which means that many previous results for single-carrier systems carry over to \OFDM systems.  
	
	By oversampling the received signal, it is possible that a better performance can be obtained than the one established by the achievable rate derived in this paper.  Future research on massive \MIMO with coarse quantization should focus on receivers that oversample the signal.
	
	\appendices
	\section{Proof of Lemma~\ref{lem:rho_and_error_var}}\label{app:rho_and_error_var}
			From \eqref{eq:expression_scaling_parameter}, the scaling factor is given by
			\begin{align}
			\bar{P}_\text{rx} \rho &= \Exp \bigl[ y_m^*[n] q_m[n] \bigr]\\
			&= \frac{1}{\sqrt{2}} \Exp\Bigl[ \bigl( \Re (y_m[n]) - j \Im (y_m[n]) \bigr) \notag\\
			&= \frac{1}{\sqrt{2}} \Exp \Bigl[ \bigl|\Re (y_m[n]) \bigr| + \bigl|\Im (y_m[n]) \bigr| \notag\\
			&\quad+ j \bigl(\Re (y_m[n]) \sign( \Im (y_m[n]) ) \notag\\
			&\quad- \Im (y_m[n]) \sign( \Re (y_m[n]) ) \bigr) \Bigr]\label{eq:step817273}\\
			&= \sqrt{2} \Exp \Bigl[ \Exp \Bigl[ \bigl| \Re (y_m[n]) \bigr| \Bigm| \{x_k[n]\} \Bigr] \Bigr]\label{eq:98298374}\\
			&= \Exp\biggl[\sqrt{\frac{2}{\pi} P_\text{rx}[n]} \biggr].\label{eq:proof_linearization_factor_step_five}
			\end{align}
			In \eqref{eq:step817273}, the imaginary part of the expected value is zero, because $\Re (y_m[n])$ and $\Im (y_m[n])$ are \IID and have zero mean.  Further, by conditioning on the transmit signals, the inner expectation in \eqref{eq:98298374} can be identified as the mean of a folded normal distributed random variable, which gives \eqref{eq:proof_linearization_factor_step_five}.
			
			The error variance is derived as
			\begin{align}
			\Exp \bigl[ |e_m[n]|^2 \bigr] &= \Exp \bigl[ |q_m[n] - \rho y_m[n]|^2 \bigr] \\
			&= 1 - \rho^2 \Exp \bigl[ |y_m[n]|^2 \bigr].
			\end{align}
			The limits in \eqref{eq:rho} and \eqref{eq:error_variance} follow directly from Lemma~\ref{lem:widebandMIMOapprox}.
	
	\section{Proof of Lemma~\ref{lem:qnoise_cond_correlation}}\label{app:qnoise_cond_correlation}
	Because they are functions of each other, the random variables
	$h_{mk}[\ell]$, $y_m[n]$, $e_m[n]$ form a Markov chain in that order.  Therefore:
	\begin{align}
	&\Exp \bigl[ h^*_{mk}[\ell] e_m[n] \bigm| \{x_k[n]\} \bigr] \notag\\
	&= \Exp \Bigl[ \Exp \bigl[ h^*_{mk}[\ell] e_m[n] \bigm| y_m [n] \bigr] \Bigm| \{x_k[n]\} \Bigr]\\
	&= \Exp \Bigl[ \Exp \bigl[ h^*_{mk}[\ell] \bigm| y_m [n] \bigr] \Exp \bigl[ e_m[n] \mid y_m [n] \bigr] \Bigm| \{x_k[n]\} \Bigr]\\
	&= \frac{x_k[n-\ell] p[\ell]}{P_\text{rx}[n]} \Exp \bigl[ y^*_m[n] (q_m[n] - \rho y_m[n]) \bigm| \{x_k[n]\} \bigr] \label{eq:constructive_combination_proof_step_five}\\
	&= \frac{x_k[n-\ell] p[\ell]}{P_\text{rx}[n]} \bigl( \Exp \bigl[ y^*_m[n] q_m[n] \bigm| \{x_k[n]\} \bigr] - \rho P_\text{rx}[n] \bigr)\\
	&= x_k[n-\ell] p[\ell] \Biggl(\frac{\sqrt{\frac{2}{\pi} P_\text{rx}[n]}}{P_\text{rx}[n]} - \frac{\Exp \Bigl[ \sqrt{\frac{2}{\pi} P_\text{rx}[n]} \Bigr]}{\bar{P}_\text{rx}} \Biggr).\label{eq:23084572908}
	\end{align}
	In \eqref{eq:constructive_combination_proof_step_five}, we used the fact that the mean of a Gaussian variable conditioned on a Gaussian-noisy observation is the \LMMSE estimate of that variable, i.e.,
	\begin{align}
	\Exp \bigl[ h_{mk}[\ell] \bigm| y_m [n], \{x_k[n]\} \bigr] = \frac{x^*_k[n-\ell] p[\ell]}{P_\text{rx}[n]} y_m[n].
	\end{align}
	In the last step \eqref{eq:23084572908}, we used the expression in \eqref{eq:expression_scaling_parameter} for $\rho$.  It can now be seen that, when $L \to \infty$, $P_\text{rx}[n] \xrightarrow{\text{a.s.}} \bar{P}_\text{rx}$ and the correlation goes to zero.

\section{Proof of Theorem~\ref{the:achievable_rate}}\label{app:achievable_rate}
	We have seen how the estimated signal can be written as the sum of the following terms:
	\begin{align}
	\freq{\hat{x}}_k[\nu] &= \rho \sum_{k'=1}^{K} \sqrt{c_{k'} \beta_{k'} P_{k'}} \big( \alpha_{kk'} \freq{x}_k[\nu] + \freq{i}_{kk'}[\nu] \big) + \rho \freq{u}'_k[\nu]\notag\\
	&\quad+ \rho \freq{z}'_k[\nu] + \freq{e}'_k[\nu].
	\end{align}
	It can be shown that each term in this sum is uncorrelated to the other terms.  
	Most correlations are easy to show, except the correlation between the error due to quantization $\freq{e}'_k[\nu]$ and the transmit signal $\freq{x}_k[\nu]$.  To show that this correlation is zero, we show that all the time-domain signals $\{e'_k[n]\}$ and $\{x_k[n]\}$ are pairwise uncorrelated if $x_k[n]$ is Gaussian.  The procedure is similar to the proof of Lemma~\ref{lem:qnoise_cond_correlation}.  We note that $x_k[n']$, $y_m[n]$, $e_m[n]$ form a Markov chain in that order.  Therefore:
	\begin{align}
	&\vspace{-0.5em}\Exp \bigl[ e^*_m[n] x_k[n'] \bigr]\notag\\
	&= \Exp \Bigl[ \Exp \bigl[ e^*_m[n] x_k[n'] \bigm| y_m [n] \bigr] \Bigr]\\
	&= \Exp \Bigl[ \Exp \bigl[ e^*_m[n] \mid y_m [n] \bigr] \Exp \bigl[ x_k[n'] \bigm| y_m [n] \bigr] \Bigr]\\
	&= \frac{\Exp[y_m^*[n] x_k[n']]}{\Exp[|y_m[n]|^2]} \Exp \bigl[ (q_m^*[n] - \rho^*y_m^*[n]) y_m[n] \bigr]\\
	&= \frac{\Exp[y_m^*[n] x_k[n']]}{\Exp[|y_m[n]|^2]} \Bigl(\Exp \bigl[ q_m^*[n] y_m[n] \bigr] - \rho^* \Exp \bigl[ |y_m[n]|^2 \bigr] \Bigr)\\
	&= 0,
	\end{align}
	for all $n$ and $n'$.  In the last step, we used \eqref{eq:expression_scaling_parameter}.
	
	The variances of $\freq{u}'_k[\nu]$ and $\freq{z}'_k[\nu]$ are given by
	\begin{align}
	\Exp\bigl[ |\freq{u}'_k[\nu]|^2 \bigr] &=  \Exp\bigl[ |\freq{u}_m[\nu]|^2 \bigr] =  \sum_{k'=1}^{K} \beta_{k'} P_{k'} \bigl(1 - c_{k'} \bigr),\\
	\Exp\bigl[ |\freq{z}'_k[\nu]|^2 \bigr] &=  \Exp\bigl[ |\freq{z}_m[\nu]|^2 \bigr] =  N_0,
	\end{align}
	By evaluating the expectations in the rate expression in \eqref{eq:raw_achievable_rate}, we obtain
	\begin{align}
	&\bigl| \Exp\bigl[ \freq{x}^*_k[\nu] \freq{\hat{x}}_k[\nu] \bigr] \bigr|^2 \to  \rho^2 c_k \beta_k P_k G_k,\\
	&\Exp\bigl[ |\freq{\hat{x}}_k[\nu]|^2 \bigr] \to   \rho^2 \Bigl( c_k \beta_k P_k G_k \notag\\
	&\quad+\sum\limits_{\mathclap{\mathclap{k'=1}}}^{K} \bigl( c_k \beta_k P_k I_{kk'} + \beta_{k'} P_{k'} (1 {-} c_{k'}) \bigr) + N_0 + Q' \Bigr),
	\end{align}
	as $L \to \infty$.  Here we used Corollary~\ref{the:noise_power}.  Letting the number of channel taps $L \to \infty$ thus gives the rate $R'_k = \log_2(1 +  \SINR_k)$, where
	\begin{align}\label{eq:938598881}
	\SINR_k = \frac{c_k \beta_k P_k G_k}{\sum_{k'=1}^{K} \big( c_k \beta_k P_k I_{kk'} {+} \beta_{k'} P_{k'} (1 {-} c_{k'}) \big) {+} N_0 {+} Q'}.
	\end{align}

	\ifCLASSOPTIONcaptionsoff
	\newpage
	\fi

	\bibliographystyle{IEEEtran}
	\bibliography{bib_forkort_namn,bibliografi}
	
	\begin{IEEEbiography}[{\includegraphics[height=1.25in, clip, keepaspectratio]{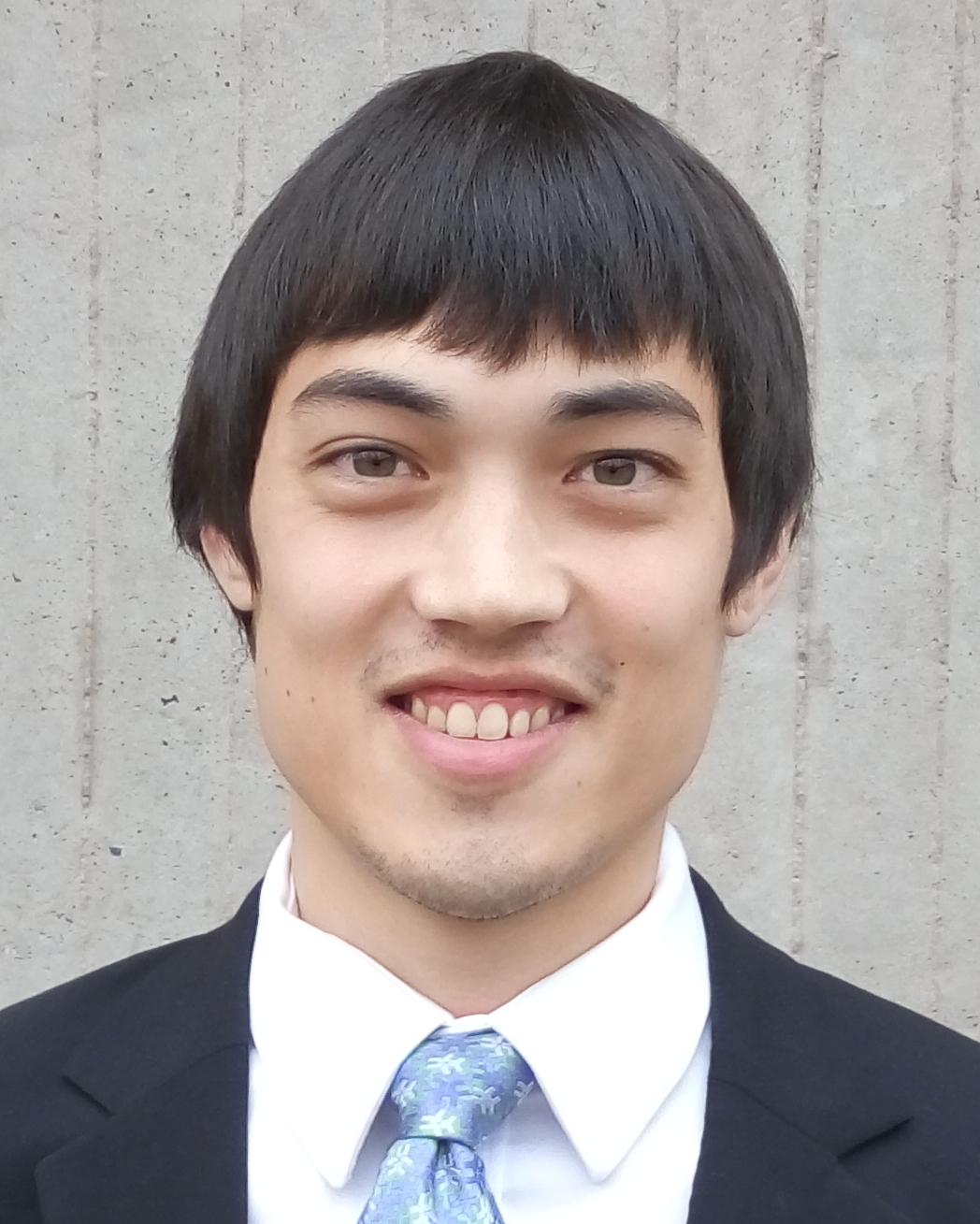}}]{Christopher Mollén}
		received the M.Sc.\ degree in 2013 and the \textit{teknologie licentiat} (Licentiate of Engineering) degree in 2016 from Linköping University, Sweden, where he is currently pursuing the Ph.D.\ degree with the Department of Electrical Engineering, Division for Communication Systems.  His research interest is low-complexity hardware implementations of massive \MIMO base stations, including low-\PAR precoding, low-resolution \ADCs, and nonlinear amplifiers.  Prior to his Ph.D.\ studies, he has worked as intern at Ericsson in Kista, Sweden, and in Shanghai, China.  From 2011 to 2012, he studied at the Eidgenössische Technische Hochschule (ETH) Zürich, Switzerland, as an exchange student in the Erasmus Programme.  And from 2015 to 2016, he visited the University of Texas at Austin as a Fulbright Scholar.
	\end{IEEEbiography}
	
	\begin{IEEEbiography}[{\includegraphics[width=1in,height=1.25in,clip,keepaspectratio]{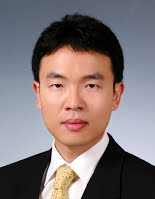}}]{Junil Choi}
		received the B.S. (with honors) and M.S. degrees in electrical engineering from Seoul National University in 2005 and 2007, respectively, and received the Ph.D. degree in electrical and computer engineering from Purdue University in 2015. He is now with the department of electrical engineering at POSTECH as an assistant professor.
		
		From 2007 to 2011, he was a member of technical staff at Samsung Advanced Institute of Technology (SAIT) and Samsung Electronics Co. Ltd. in Korea, where he contributed to advanced codebook and feedback framework designs for the 3GPP LTE/LTE-Advanced and IEEE 802.16m standards. Before joining POSTECH, he was a postdoctoral fellow at The University of Texas at Austin. His research interests are in the design and analysis of massive \MIMO, mmWave communication systems, distributed reception, and vehicular communication systems.
		
		Dr. Choi was a co-recipient of a 2015 IEEE Signal Processing Society Best Paper Award, the 2013 Global Communications Conference (GLOBECOM) Signal Processing for Communications Symposium Best Paper Award and a 2008 Global Samsung Technical Conference best paper award. He was awarded the Michael and Katherine Birck Fellowship from Purdue University in 2011; the Korean Government Scholarship Program for Study Overseas in 2011-2013; the Purdue University ECE Graduate Student Association (GSA) Outstanding Graduate Student Award in 2013; and the Purdue College of Engineering Outstanding Student Research Award in 2014. 
	\end{IEEEbiography}
	
	\begin{IEEEbiography}[{\includegraphics[width=1in,height=1.25in,clip,keepaspectratio]{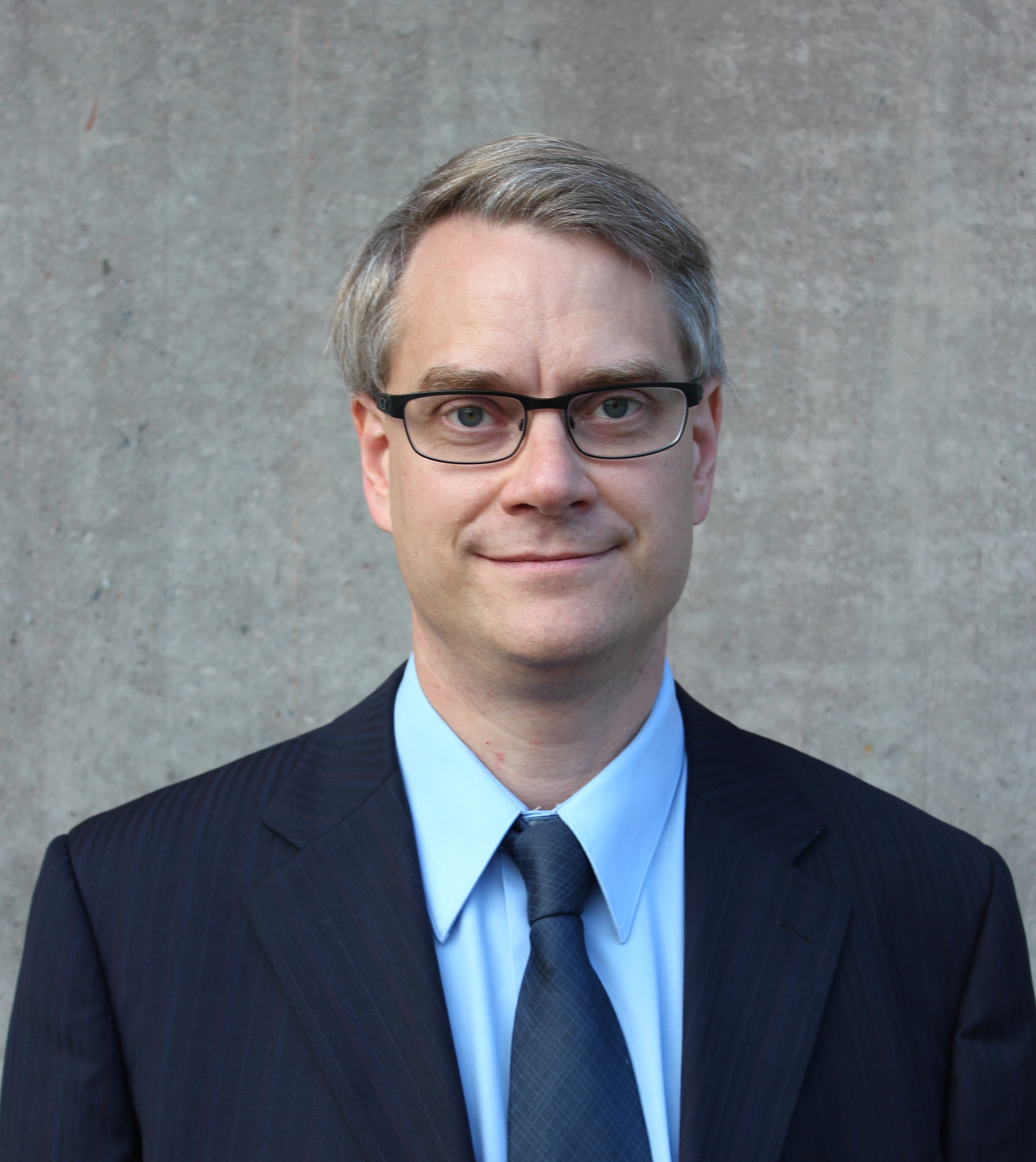}}]{Erik G. Larsson}
		is Professor of Communication Systems at Link\"oping University (LiU) in Link\"oping, Sweden. He 
		previously worked for the Royal Institute of Technology (KTH) in Stockholm, Sweden, the University of Florida, USA, the George Washington University, USA, and Ericsson Research, Sweden.  In 2015 he was
		a Visiting Fellow at Princeton University, USA, for four months. He received his Ph.D. degree from Uppsala University, Sweden, in 2002.  
		
		His main professional interests are within the areas of wireless communications and signal processing. He has co-authored some 130 journal papers on these topics, he is co-author of the two Cambridge University Press textbooks \emph{Space-Time Block Coding for Wireless Communications} (2003) and \emph{Fundamentals of Massive MIMO} (2016).  He is co-inventor on 16 issued and many pending patents on wireless technology.
		
		He served as Associate Editor for, among others, the \emph{IEEE Transactions on Communications} (2010-2014) and \emph{IEEE Transactions on Signal Processing} (2006-2010).  He serves as  chair of the IEEE Signal Processing Society SPCOM technical committee in 2015--2016 and he served as chair of the steering committee for the \emph{IEEE Wireless Communications Letters} in 2014--2015.  He was the General Chair of the Asilomar Conference on Signals, Systems and Computers in 2015, and Technical Chair in 2012.  
		
		He received the \emph{IEEE Signal Processing Magazine} Best Column Award twice, in 2012 and 2014, and the IEEE ComSoc Stephen O. Rice Prize in Communications Theory in 2015.  He is a Fellow of the IEEE.
	\end{IEEEbiography}
	
	\begin{IEEEbiography}[{\includegraphics[width=1in,height=1.25in,clip,keepaspectratio]{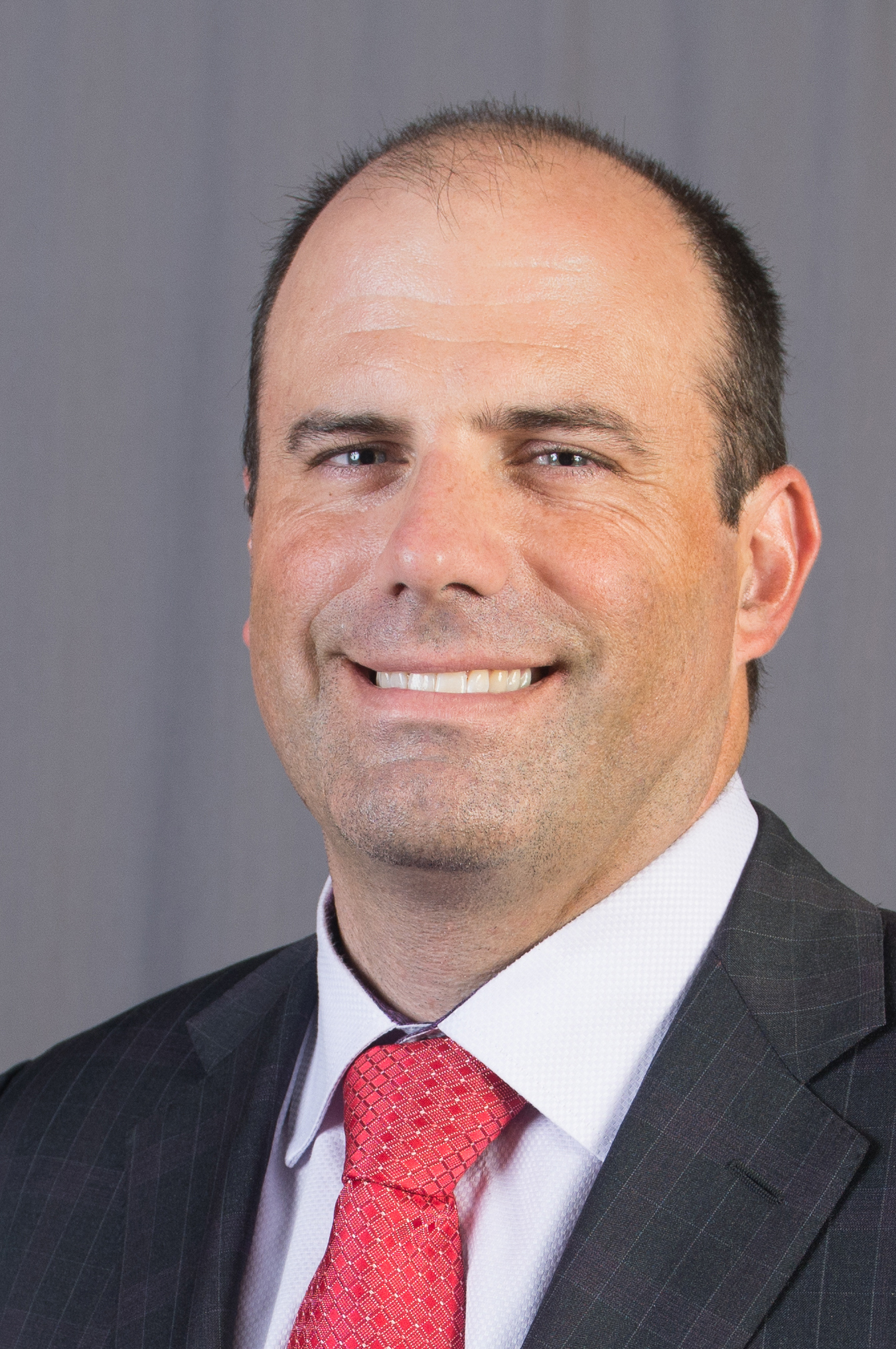}}]{Robert W. Heath, Jr.}
		(S'96 - M'01 - SM'06 - F'11)  received the B.S. and M.S. degrees from the University of Virginia, Charlottesville, VA, in 1996 and 1997 respectively, and the Ph.D. from Stanford University, Stanford, CA, in 2002, all in electrical engineering. From 1998 to 2001, he was a Senior Member of the Technical Staff then a Senior Consultant at Iospan Wireless Inc, San Jose, CA where he worked on the design and implementation of the physical and link layers of the first commercial \MIMO-\OFDM communication system. Since January 2002, he has been with the Department of Electrical and Computer Engineering at The University of Texas at Austin where he is a Cullen Trust for Higher Education Endowed Professor, and is a Member of the Wireless Networking and Communications Group. He is also President and CEO of MIMO Wireless Inc. He is a co-author of the book Millimeter Wave Wireless Communications published by Prentice Hall in 2014 and author of Digital Wireless Communication: Physical Layer Exploration Lab Using the NI USRP published by the National Technology and Science Press in 2012. 
		
		Dr.\ Heath has been a co-author of several best paper award recipients including recently the 2010 and 2013 EURASIP Journal on Wireless Communications and Networking best paper awards, the 2012 Signal Processing Magazine best paper award, a 2013 Signal Processing Society best paper award, 2014 EURASIP Journal on Advances in Signal Processing best paper award, the 2014 Journal of Communications and Networks best paper award, the 2016 IEEE Communications Society Fred W. Ellersick Prize, and the 2016 IEEE Communications and  Information Theory Societies Joint Paper Award. He was a distinguished lecturer in the IEEE Signal Processing Society and is an ISI Highly Cited Researcher. He is also an elected member of the Board of Governors for the IEEE Signal Processing Society, a licensed Amateur Radio Operator, a Private Pilot, and a registered Professional Engineer in Texas. 
	\end{IEEEbiography}	
\end{document}